\documentclass[preprint2]{aastex}

\usepackage{graphicx}
\usepackage{amsmath}
\usepackage{amssymb}

\newcommand{\nthp}{N$_2$H$^+$}
\newcommand{\ntdp}{N$_2$D$^+$}
\newcommand{\Dfrac}{$D_{\rm frac}^{{\rm N}_2{\rm H}^+}$}
\newcommand{\Dfracdef}{$D_{\rm frac}^{{\rm N}_2{\rm H}^+} \equiv {\rm N}_2{\rm D}^+/{\rm N}_2{\rm H}^+$}
\newcommand{\OPRH}{OPR$^{{\rm H}_2}$}
\newcommand{\OPR}{OPR$_{0}^{{\rm H}_2}$}

\shorttitle{Deuteration in Pre-Stellar Cores}
\shortauthors{Goodson et al.}

\begin{document}

\title{Structure, Dynamics and Deuterium Fractionation of Massive Pre-Stellar Cores}
\author{Matthew D. Goodson}
\affil{Dept. of Physics and Astronomy, University of North Carolina at Chapel Hill, Chapel Hill, NC 27599-3255, USA}
\email{mgoodson@unc.edu}
\author{Shuo Kong}
\affil{Dept. of Astronomy, University of Florida, Gainesville, FL 32611, USA}
\affil{Dept. of Astronomy, Yale University, New Haven, CT 06520-8101, USA}
\author{Jonathan C. Tan}
\affil{Dept. of Astronomy, University of Florida, Gainesville, FL 32611, USA}
\affil{Dept. of Physics, University of Florida, Gainesville, FL 32611, USA}
\author{Fabian Heitsch}
\affil{Dept. of Physics and Astronomy, University of North Carolina at Chapel Hill, Chapel Hill, NC 27599-3255, USA}
\author{Paola Caselli}
\affil{Max-Planck-Institute for Extraterrestrial Physics (MPE), Giessenbachstr. 1, D-85748 Garching, Germany}

\begin{abstract}
High levels of deuterium fraction in N$_2$H$^+$ are observed in some pre-stellar cores. Single-zone chemical models find that the timescale required to reach observed values ($D_{\rm frac}^{{\rm N}_2{\rm H}^+} \equiv {\rm N}_2{\rm D}^+/{\rm N}_2{\rm H}^+ \gtrsim 0.1$) is longer than the free-fall time, possibly ten times longer. Here, we explore the deuteration of turbulent, magnetized cores with 3D magnetohydrodynamics simulations. We use an approximate chemical model to follow the growth in abundances of N$_2$H$^+$ and N$_2$D$^+$. We then examine the dynamics of the core using each tracer for comparison to observations. We find that the velocity dispersion of the core as traced by N$_2$D$^+$ appears slightly sub-virial compared to predictions of the Turbulent Core Model of McKee \& Tan, except at late times just before the onset of protostar formation. By varying the initial mass surface density, the magnetic energy, the chemical age, and the ortho-to-para ratio of H$_2$, we also determine the physical and temporal properties required for high deuteration. We find that low initial ortho-to-para ratios ($\lesssim 0.01$) and/or multiple free-fall times ($\gtrsim 3$) of prior chemical evolution are necessary to reach the observed values of deuterium fraction in pre-stellar cores.
\end{abstract}

\maketitle

\section{Introduction}\label{s:intro}
\subsection{Massive Star Formation}\label{ss:massivestars}
Massive stars play a central role in galactic evolution through feedback and metal enrichment, yet the physical processes and conditions involved in massive star formation remain uncertain \citep{2014prpl.conf..149T}. The relative rarity of massive stars and thus their typical large distances from us, along with their deeply embedded formation environments, make it difficult to observe details of the massive star formation process.

There are two main theories for massive star formation: 1) Core Accretion models, e.g., the Turbulent Core Accretion model \citep[][hereafter MT03]{2003ApJ...585..850M}, which assumes near-virialized starting conditions for relatively ordered collapse; and 2) the Competitive Accretion model \citep{2001MNRAS.323..785B}, which posits fragmentation and subsequent accretion by multiple stars from a turbulent, globally collapsing medium. Distinguishing these two scenarios relies on disentangling the numerous physical processes involved, such as turbulent motions, magnetic fields, and feedback.

Numerical modeling is one means to extricate the various processes. Previous simulations of massive star formation have focused on the role of turbulence, magnetic fields, and radiation in clump fragmentation. \citet{2011MNRAS.413.2741G} investigated the fragmentation of hydrodynamic clumps, examining the effect of the initial density profile and turbulent driving. The authors found that single massive stars are more likely to form from centrally-concentrated initial conditions, while the details of the turbulence are relatively unimportant. Numerous authors \citep{2007ApJ...656..959K,2010ApJ...713.1120K,2011ApJ...729...72P,2011ApJ...740..107C,2011ApJ...742L...9C,2013ApJ...766...97M} have demonstrated that radiative feedback from protostars inhibits fragmentation of the clump. Magnetohydrodynamics (MHD) simulations both neglecting radiation \citep{2011A&A...528A..72H,2011MNRAS.417.1054S,2012MNRAS.422..347S} and with radiation \citep{2011ApJ...729...72P,2011ApJ...742L...9C,2013ApJ...766...97M} indicate that even a weak magnetic field suppresses clump fragmentation, and increasing the field strength further reduces the fragmentation.

In all of the aforementioned MHD numerical studies, the magnetic field strength is initially super-critical, i.e., the field cannot prevent gravitational collapse. The central pre-stellar core contracts rapidly, forming a protostar within one to two free-fall times. Yet the timescale of core collapse remains an open question. In the Competitive Accretion model, cores form and rapidly collapse on the order of the free-fall time. In the Turbulent Core model, the cores persist longer -- at least one dynamical time -- possibly supported by magnetic fields and turbulence near virial balance. Indeed, some observed cores exhibit supersonic linewidths consistent with virial balance \citep{2013ApJ...779...96T,2016arXiv160906008K}. Yet, velocity dispersions due to virial equilibrium or energy equipartition (consistent with free-fall) differ only by a factor of $\sqrt{2}$ \citep{2007ApJ...657..870V}. Therefore, even a clear distinction between virial equilibrium and free-fall collapse based on velocity dispersion seems difficult. However, we note that where they have been measured, observed infall speeds generally generally seem to be small, i.e., $\sim 1/3$ of the free-fall velocity \citep{2016A&A...585A.149W}.

\subsection{Deuteration as a Chemical Clock}\label{ss:dclock}

An alternative means to probe the age and state of starless cores is using chemical tracers, in particular deuterated molecules. In sufficiently dense ($n_{\rm H} > 10^5$~cm$^{-3}$), cold ($T < 20$~K) environments, CO freeze-out opens a pathway for ion-neutral reactions that increase the deuterium fraction, i.e., the ratio of deuterated to non-deuterated species, $D_{\rm frac}$. For a full review of deuteration processes, see \citet{2014prpl.conf..859C}. Observationally, deuterated molecules are excellent probes of pre-stellar gas. \citet{2002ApJ...565..344C} traced low-mass star forming regions with N$_2$D$^+$ and DCO$^+$, finding deuterium fractions $D_{\rm frac} \gtrsim 0.1$, several orders of magnitude above the cosmic deuterium ratio (D/H $\sim 10^{-5}$). Similarly, \citet[][hereafter T13]{2013ApJ...779...96T} identified high-mass star-forming regions in infrared dark clouds (IRDCs) with the same deuterated molecules. \citet[][hereafter K16]{2016ApJ...821...94K} has subsequently estimated the deuterium fraction of N$_2$H$^+$ in these regions to be of comparable values to those in low-mass pre-stellar cores (\Dfracdef$\gtrsim 0.1$) \citep[see also][]{2011A&A...529L...7F}.

As deuteration is expected to begin only when pre-stellar core conditions are satisfied, the deuterium fraction may be a useful estimator of core age. \citet[][hereafter K15]{2015ApJ...804...98K} developed a time-dependent astrochemical network to model the evolution of deuterium-bearing molecules. The authors followed the chemistry in a single zone with fixed physical conditions or with simple density evolution. Under typical core conditions, the K15 models suggest that the deuteration process is slow, with up to ten free-fall times required to reach the observed values of \Dfrac.

Moving beyond single-zone chemical models is a difficult task, as the complex reaction network requires extensive computational resources. \citet{2013A&A...551A..38P} coupled the deuterium network of \citet{2009A&A...494..623P} with a 1D spherically-symmetric hydrodynamic calculation. The simulations followed deuteration in 200 radial zones during collapse of a low-mass pre-stellar core from a uniform, static state. In disagreement with K15, \citet{2013A&A...551A..38P} determined that fast collapse is preferred, as steady-state abundances determined from the model were typically much higher than observed. However, the models of \citet{2009A&A...494..623P} and \citet{2013A&A...551A..38P} begin with very high initial depletion factors, which greatly shortens the deuteration timescale. A full discussion and comparison is presented in K15, but it is worth noting that, given similar initial conditions, the models of K15 agree with \citet{2009A&A...494..623P} to within a factor of 3. 

If large-scale magnetic fields are present, the assumption of radial symmetry during collapse will not hold, as flux-freezing prevents significant collapse in directions perpendicular to the field. Further, the turbulent motions within the core are not fully captured in 1D simulations. Indeed, the chemical evolution may be altered by non-linear effects such as density fluctuations and turbulent diffusion. Implementing a full chemical network in high-resolution 3D simulations is currently not feasible given computational limits. One option may be to reduce the number of reactions and reactants; however, this would negatively affect the accuracy of the chemistry. Here, we develop an alternative approach.

We construct an approximate deuterium chemistry model built on the full astrochemical network results of K15. By parameterizing the results across a wide range of densities, we formulate a robust and efficient method to follow the growth and deuteration of \nthp{} in 3D MHD simulations of massive core collapse. We generate a turbulent, magnetized pre-stellar core according to the paradigm of MT03, and we model the collapse of the core until the first protostar forms. We simultaneously follow the chemical evolution of \nthp{} and \ntdp{} and compare to observed massive pre-stellar cores. By varying the initial conditions, such as the mass surface density, magnetic energy, chemical age, and initial ortho-to-para ratio of H$_2$, we can estimate the core properties necessary to match observed deuterium abundances. 

We observe in our simulations that the collapse occurs on roughly the free-fall time, regardless of the initial mass surface density or magnetic field strength. We conclude that reaching the observed deuterium fractions requires significant prior chemical evolution, low initial ortho-to-para ratio, and/or slower collapse, possibly by stronger magnetic fields or sustained turbulence.

We outline our numerical methods, including initial conditions and chemical model in (\S\ref{s:methods}). The results of our simulations are presented and discussed in \S\ref{s:results}. We discuss the implications for massive star formation in \S\ref{s:discussion} before concluding in \S\ref{s:conclusions}.

\section{Methods}\label{s:methods}

We use a modified version of \textsc{Athena} \citep{2008ApJS..178..137S} version 4.2 to solve the equations of ideal, inviscid MHD:
\begin{eqnarray}\label{eq:mhd}
\frac{\partial \rho}{\partial t} + \nabla \cdot (\rho {\bf u}) &=& 0 \\
\frac{\partial\rho {\bf u}}{\partial t} + \nabla \cdot (\rho {\bf u u} - {\bf B B} + \frac{{\bf B} \cdot {\bf B}}{2} + P{\bf I}) &=&  0\\
\frac{\partial E}{\partial t} + \nabla \cdot [(E+P+\frac{{\bf B} \cdot {\bf B}}{2}) \bf{u}] &=& 0 \\
\frac{\partial {\bf B}}{\partial t} - \nabla \times ({\bf u} \times {\bf B}) &=& 0
\end{eqnarray} with the density $\rho$, the velocity vector $\bf{u}$, the magnetic field vector ${\bf B}$, the thermal pressure $P$, the unit dyad {\bf I}, and the total energy density $E$:
\begin{equation}\label{eq:eos}
E = \frac{P}{\gamma -1} + \frac{1}{2} \rho |\mathbf{u}|^2 + \frac{{\bf B} \cdot {\bf B}}{2}.
\end{equation}We use a passive color field $C$ to trace core material:
\begin{equation}\label{eq:color}
\frac{\partial \rho C}{\partial t}  +  \nabla \cdot (\rho {\bf u} C) = 0.
\end{equation}We also evolve several scalar fields to trace the chemistry:
\begin{equation}\label{eq:passive}
\frac{\partial \rho [X]}{\partial t}  +  \nabla \cdot (\rho {\bf u} [X]) = S([X])
\end{equation}with the fractional abundance $[X]$ for some species $X$ relative to hydrogen, and a source term $S$. Full details of the chemical model are presented in \S\ref{ss:chemistry}.

We use the directionally unsplit van Leer (VL) integrator \citep{2009NewA...14..139S} with second order reconstruction in the primitive variables \citep{1984JCoPh..54..174C} and the HLLD Riemann solver \citep{Toro2009}. Simulations are performed on Cartesian grids in three dimensions. To obtain an approximately isothermal equation of state, we set the ratio of specific heats $\gamma = C_P/C_V = 1.001$. We do not include radiation pressure or feedback; we do include self-gravity.

\subsection{Setup and Initial Conditions}\label{ss:setup}

We initialize a spherical core according to the relations of MT03. We set the core mass $M_{\rm c} = 60~$M$_\odot$ and the density power law exponent $k_\rho = 1.5$. For our fiducial core, we set the clump mass surface density $\Sigma_{\rm cl} = 0.3 $~g~cm$^{-2}$, consistent with the estimates of T13 observed cores. With these values, MT03 prescribes the radius of the core
\begin{equation}
R_{\rm c} = 0.057~\left(\frac{\Sigma_{\rm cl}}{1~{\rm g~cm}^{-2}}\right)^{-1/2}~{\rm pc}~\to~0.10~{\rm pc}
\end{equation}and the number density of hydrogen at the surface 
\begin{multline}
n_{{\rm H},s} = 1.16 \times 10^6~\left(\frac{\Sigma_{\rm cl}}{1~{\rm g~cm}^{-2}}\right)^{3/2}~{\rm cm}^{-3} \\ \to~1.82 \times 10^5~{\rm cm}^{-3},
\end{multline}where the value after the arrow is for the fiducial model. The mean number density in the core is $\bar{n} = 1.97 \times 10^5$~cm$^{-3}$, and the average free-fall time of the core is
\begin{equation}
t_{\rm ff} = \sqrt{\frac{3 \pi}{32 G \bar{\rho}}} ~\to~ 76 ~ {\rm kyr},
\end{equation} with average density $\bar{\rho} = \bar{n} \mu m_{\rm H}$, the mean molecular weight $\mu = 2.33$, and the mass of hydrogen $m_{\rm H}$. To trace the core, we initialize the passive color field $C$ to unity for $r \le R_{\rm c}$ and zero otherwise. We also perform simulations with lower initial mass surface density of the surrounding clump ($\Sigma_{\rm cl} = 0.1 $~g~cm$^{-2}$); the relevant parameters for both cases are summarized in Table \ref{tab:sims}.

\floattable
\begin{deluxetable}{ccccccccccc}
\tablecaption{Summary of Simulations \label{tab:sims}}
\tablehead{ \colhead{} & \colhead{$\Sigma_{\rm cl}$} & \colhead{$\mu_{\Phi}$} & \colhead{$M_{\rm c}$} & \colhead{$R_{\rm c}$} & \colhead{$\bar{n}$} & \colhead{$t_{\rm ff}$} & \colhead{$B_{\rm c}$} & \colhead{$\alpha$} & \colhead{$\sigma$} & \colhead{$\delta$} \\ \colhead{Run Name} & \colhead{(g~cm$^{-2}$)} & \colhead{} & \colhead{(M$_\sun$)} & \colhead{(pc)} & \colhead{(cm$^{-3}$)} & \colhead{(kyr)} & \colhead{(mG)} & \colhead{} & \colhead{(km~s$^{-1}$)} & \colhead{(AU)} }
\startdata
S3M2 & 0.3 & 2 & 60 & 0.104 & 1.97$\times10^5$ &  76 & 0.803 & 2 & 0.99 & 213 \\
S1M2 & 0.1 & 2 & 60 & 0.180 & 3.79$\times10^4$ & 173 & 0.268 & 2 & 0.76 & 365 \\
S3M1 & 0.3 & 1 & 60 & 0.104 & 1.97$\times10^5$ &  76 & 1.606 & 2 & 0.99 & 213 \\
S1M1 & 0.1 & 1 & 60 & 0.180 & 3.79$\times10^4$ & 173 & 0.536 & 2 & 0.76 & 365 \\
\enddata
\tablecomments{Fiducial simulation is S3M2.}
\end{deluxetable}

\subsubsection{Density Structure}\label{sss:denssetup}

The core has a density profile $\rho(r) \propto r^{-1.5}$, which is consistent with observations of massive pre-stellar cores \citep{2012ApJ...754....5B}. We set the core to a constant temperature $T_{\rm c} = 15$~K; thus the thermal pressure in the core follows the same power law as the density. The sound speed in the core $c_s = \sqrt{k_B T/(\mu m_{\rm H})} = 0.2$~km~s$^{-1}$, with the Boltzmann constant $k_B$. To prevent divergence as $r \to 0$, we flatten the profile over an inner radius, $R_{\rm f} = 0.15 R_{\rm c}$. We calculate the central density $n_{\rm c} =  n_{\rm s}[1.0 + (R_{\rm c}/R_{\rm f})^{k_\rho}] \to 1.99 \times 10^6$~cm$^{-3}$. We impose an order of magnitude jump in the density at the core surface, which is smoothed by a hyperbolic tangent profile with $R_{\rm s} = 0.05 R_{\rm c}$. The density in the ambient medium is constant at $n_0 = 0.1 n_{\rm s}$. The overall density profile is given by
\begin{equation}
n(r) = n_0 + \frac{n_{\rm c} - n_0}{1 + (r/R_{\rm f})^{k_\rho}}(0.5 - 0.5 \tanh{[\frac{r-R_{\rm c}}{R_{\rm s}}]}).
\end{equation}The ambient temperature $T_0$ is determined by thermal pressure balance with the core: $T_0 = 10 T_{\rm c} = 150$~K. This mimics the effective pressure of the surrounding clump medium, which is expected to be dominated by non-thermal mechanisms, e.g., turbulence. The actual temperature of the clump is expected to be quite similar to that of the core. 

The core is centered in a cubic simulation box of side length $L = 5 R_{\rm c} \to 0.52$~pc with spatial resolution $\delta = L/512 \to 213$~AU. We use periodic boundary conditions to prevent gravitational evacuation at the box edges; the core is sufficiently padded to prevent any interactions with the boundaries. We use a periodic FFT solver to calculate the gravitational potential. Eventually runaway collapse in a few cells drives the global time-step to nearly zero. The collapse could be followed longer by the addition of sink particles, but as we are only interested in pre-stellar conditions, we terminate the simulation at this point.

\begin{figure}
  \includegraphics[width=\columnwidth]{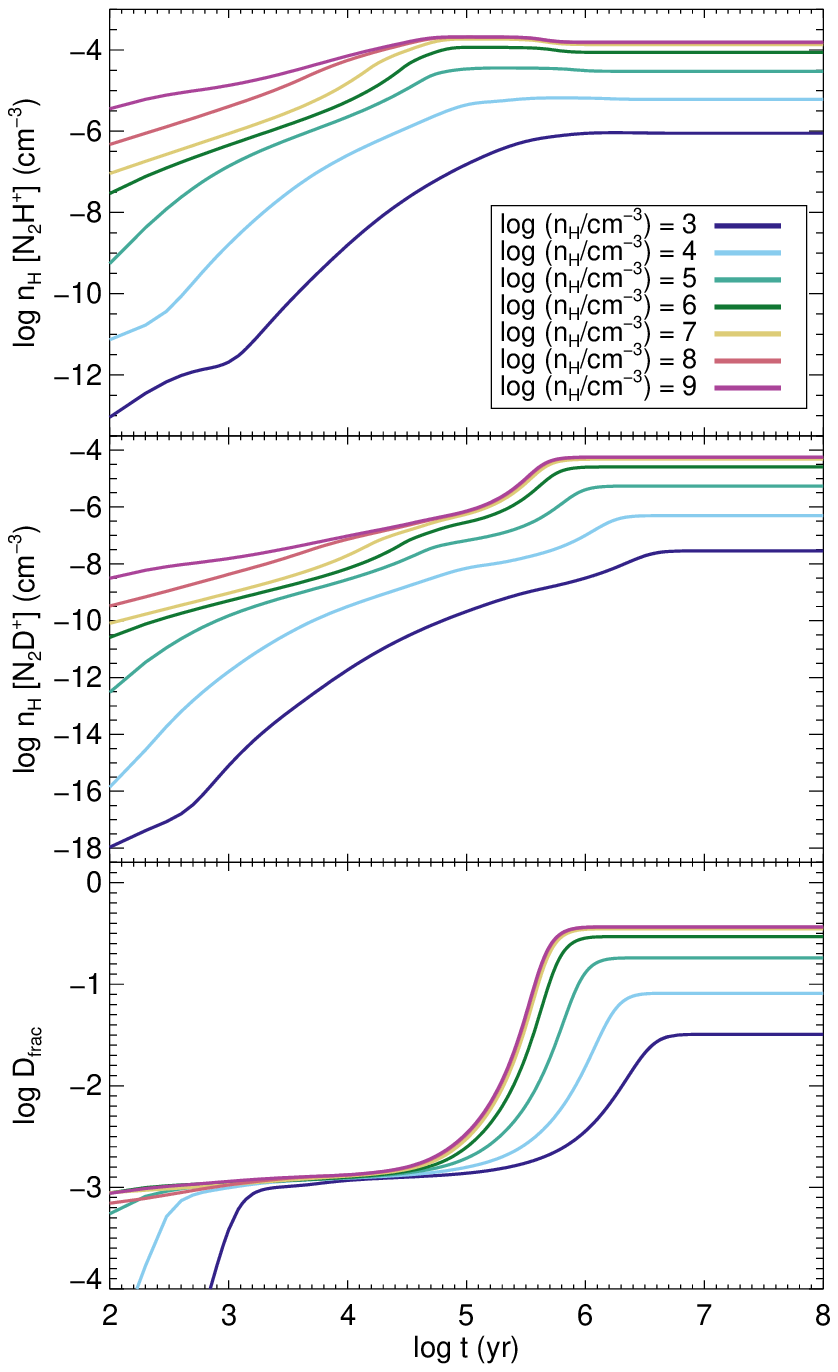}
  \caption{Time evolution of chemical number density and deuterium fraction from K15 for various hydrogen number densities $n_{\rm H}$. The number density is computed as $n_{\rm H} [X]$, where $[X]$ is the relative abundance of species $X$. From top to bottom, the number density of \nthp, of \ntdp, and the deuterium fraction, \Dfracdef. Results are obtained with K15 fiducial parameters except that \OPR~=~0.1. The time required to reach equilibrium decreases with increasing density. Additionally, the equilibrium values for both species abundance and the deuterium fraction increase with increasing density, reaching \Dfrac{} $\gtrsim 0.3$ at $n_{\rm H} = 10^9~{\rm cm}^{-3}$.}
  \label{f:denscompare}
\end{figure}

\subsubsection{Magnetic Fields}\label{sss:magsetup}

We initialize a cylindrically-symmetric magnetic field in the $z$-direction, similar to the field geometry of \citet{2013ApJ...766...97M}. The field strength is determined by the desired mass-to-flux ratio normalized to the critical value \citep{1976ApJ...210..326M}:
\begin{equation}
\mu_\Phi = \frac{M}{M_\Phi} = \frac{2 \pi G^{1/2} M}{\Phi},
\end{equation} where $M_\Phi$ is the critical mass-to-flux value and $\Phi$ is the magnetic flux through the center of the core. To maintain approximately constant $\mu_\Phi$ throughout the core, the field strength decreases as $r^{-0.5}$; then the magnetic pressure $B^2/(8 \pi) \propto r^{-1}$, as in MT03. For a given $\mu_\Phi$, we calculate the field strength at the surface of the core $B_{\rm s}$:
\begin{equation}
B_{\rm s} = \frac{3}{2} \frac{G^{1/2} M_{\rm c}}{\mu_\Phi R_{\rm c}^2} \to 0.22~{\rm mG}.
\end{equation}Similar to our treatment of the density, we smooth the magnetic field profile both at the center of the core and at the edge of the core. The field in the ambient medium is uniform at $B_0 = B_{\rm s}$, and the overall magnetic field profile is given by
\begin{equation}
B(\xi) = B_0 + \frac{B_{\rm c} - B_0}{1 + (\xi/R_{\rm f})^{0.5}}(0.5 - 0.5 \tanh{[\frac{\xi-R_{\rm c}}{R_{\rm s}}]}),
\end{equation}where $\xi \equiv \sqrt{x^2 + y^2}$ is the distance from the $z$-axis and $B_{\rm c}$ is the central field strength, given by $B_{\rm c} =  B_{\rm s}[1.0 + (R_{\rm c}/R_{\rm f})^{0.5}]$. Our fiducial simulation uses a slightly super-critical mass-to-flux ratio ($\mu_\Phi = 2$), in accord with observations of dense molecular gas \citep{2012ARA&A..50...29C}; then the central field strength $B_{\rm c} \to 0.80$~mG. We also perform simulations with a stronger magnetic field, corresponding to critical mass-to-flux ratio ($\mu_\Phi = 1$). Relevant parameters in both cases are summarized in Table \ref{tab:sims}.

\subsubsection{Turbulence}\label{sss:turbsetup}

We initialize supersonic turbulence in the cores with random velocity perturbations. The turbulence is generated in a method similar to that described in \citet{MacLow1999}: amplitudes are drawn from a random Gaussian with a Fourier power spectrum of form $\left| \delta \mathbf{v}_k \right| \propto k^{-2}$, with $1.0 < k L / 2 \pi  < N/2 $, where $k$ is the wavenumber, $L$ is the box size, and $N$ is the number of cells. We apply fully solenoidal (divergence-free) perturbations. The initial perturbation has a one-dimensional velocity dispersion $\sigma$ calculated from the virial relation, $\alpha \equiv 5 \sigma^2  R_{\rm c} / (G M_{\rm c})$ \citep{1992ApJ...395..140B}. Because we do not initialize any density perturbations, we set the core to be initially super-virial ($\alpha = 2$); the initial velocity dispersion in the fiducial simulation is then $\sigma \to 0.99$~km~s$^{-1}$. This value is close to the velocity dispersion of a virialized core including external pressure terms, given in T13:
\begin{multline}\label{eq:sigmacvir}
\sigma_{\rm c,vir} = 1.09 \left(\frac{M_{c}}{60~{\rm M}_\sun}\right)^{1/4} \left(\frac{\Sigma_{\rm cl}}{1~{\rm g cm}^{-2}}\right)^{1/4} {\rm km~s}^{-1} \\ \to 0.80~{\rm km~s}^{-1}.
\end{multline}We do not drive the turbulence; energy is only injected at initialization.

\subsection{Chemistry}\label{ss:chemistry}

We follow the evolution of two molecular species in our simulations: N$_2$H$^+$ and N$_2$D$^+$. The fractional abundance of each species is advected with the fluid as a passive color field (Eq. \ref{eq:passive}). We use an approximate chemical model based on the results of K15, in which the authors presented a time-dependent chemical network for the evolution of \nthp{} and \ntdp{} in a single-zone approximation. We combine results from across the K15 parameter space into a unified model to predict the initial chemical abundances and growth rates.

In K15, the authors examined the influence of numerous physical conditions and found that the results depend strongly on the number density of hydrogen, $n_{\rm H}$, and the initial ortho-to-para ratio of H$_2$, \OPR. Deuteration is most efficient when the number density is high and \OPRH{} is low. Unfortunately, \OPRH{} is not easy to estimate from observations. The statistical expectation for \OPRH{} at H$_2$ formation on grains is \OPRH~=~3.0; \OPRH{} then decreases as ortho-H$_2$ is destroyed. We test the effect of different initial \OPRH{} values by including three sets of K15 simulations: \OPR~=~1.0, 0.1, and 0.01. 

For a given \OPR, we use a suite of 55 uniform density models from K15 to construct our approximate model, spanning hydrogen number densities from $10^3$ to $10^9$~cm$^{-3}$. All models use the fiducial parameters of K15: gas temperature $T = 15~{\rm K}$, cosmic ray ionization rate $\zeta = 2.5\times 10^{-17}~{\rm s}^{-1}$, heavy-element depletion factor $f_{\rm D} = 10$, radiation field (relative to Habing field) $G_0 = 1$, and visual extinction $A_V = 30~{\rm mag}$. We note that, while $\zeta \approx 3 \times 10^{-16}~{\rm s}^{-1}$ in diffuse gas \citep{2012ApJ...745...91I}, cosmic rays are attenuated in dense starless cores to a value approximately an order of magnitude lower \citep{2009A&A...501..619P,2010MNRAS.402.1625K10}. We also note that for these conditions of high extinction the radiation field plays a negligible role. Each K15 model provides the time evolution of the fractional abundance of species X, denoted $[X](t)$, over 100~Myr. Figure \ref{f:denscompare} presents the fiducial results of K15 for varying hydrogen number density $n_{\rm H}$ at \OPR~=~0.1, with the deuterium fraction \Dfracdef. 

\subsubsection{Chemical Age}\label{sss:chemage}

To set the initial condition for the molecular abundances, we must make assumptions about the previous history of the gas. Deuteration begins as CO starts to freeze out, which occurred prior to $t=0$ for our simulation; the exact amount of prior time is unknown. We therefore investigate four chemical starting times, $t_{\rm chem}$, which for simplicity we make multiples of the mean core free-fall time: $t_{\rm chem}$~=~0, 1, 3, and 10~$t_{\rm ff}$. For $t_{\rm chem} =0$, we assume [\nthp]=[\ntdp]=0.0. For all other $t_{\rm chem}$, we reference the constant density runs of K15. We first interpolate the K15 results using a cubic spline onto an $n_{\rm H}$-$t$ grid of $1000^2$ support points. This finer grid then functions as a look-up table; given a particular starting time $t_{\rm chem}$ and density $n_{\rm H}$, we estimate the chemical abundances using bi-linear interpolation. This method implicitly assumes that the gas has been in its current configuration for the duration of $t_{\rm chem}$. While this is clearly an idealization, it provides a simple test of the importance of the previous history of the gas.

\subsubsection{Chemical Growth Rates}\label{sss:chemgrowth}

\begin{figure}
  \includegraphics[width=\columnwidth]{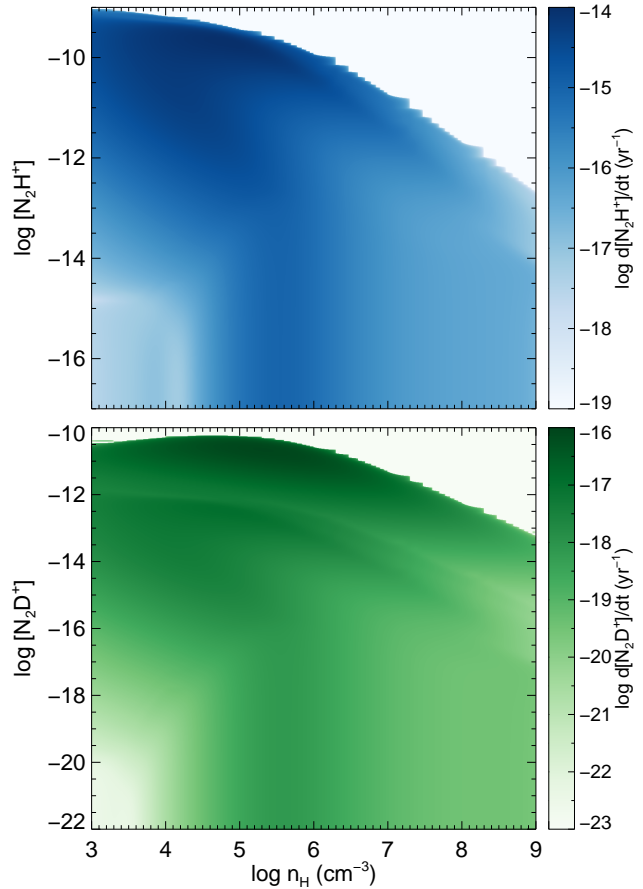}
  \caption{Example look-up grid for the chemical evolution of (top) [\nthp] and (bottom) [\ntdp] in our simulation. The grid shown is for \OPR~=~0.1 and is generated from the chemical network modeling of K15. A growth rate d$[X]/$d$t$ is estimated for each species $X$ using bi-linear interpolation based on the current hydrogen number density $n_{\rm H}$ and relative abundance $[X]$.}
  \label{f:interpgridnx}
\end{figure}

The time evolution of each chemical species at a given density is provided by K15. From these runs, we can calculate a growth rate, d$[X]/$d$t$ as a function of time and density. If our simulations maintained a constant density, we could use the absolute time to determine the growth rate and easily evolve the abundances. However, in a dynamical simulation with non-linear density evolution, the time dependence is not straightforward. If we restrict the abundances to strictly grow monotonically, we can parameterize the chemical growth rate as a function of the chemical abundance itself and remove the time dependence. In the chemical modeling results of K15, [\ntdp] strictly monotonically increases, and [\nthp] monotonically increases except for a slight decrease very near equilibrium. As the effect is relatively small ($\lesssim 30\%$), we ignore any decreases in chemical abundances. With this modification, we can parameterize the growth rate as a function of the current species abundance. 

We calculate the time derivative as a function of chemical abundance for each of the constant-density runs performed in K15 using a second-order central difference. For computational efficiency, we then interpolate the results onto a $1000^2$ $n_{\rm H}$-$[X]$ look-up grid. Figure \ref{f:interpgridnx} shows an example grid for \OPR~=~0.1. For each cell and at each time step in the simulation, the growth rate is estimated by bi-linear interpolation based on the current density and fractional abundance. The total source term $S([X]) = \rho~ dt~(d[X]/dt)$ is calculated using a sub-cycled fourth-order Runge-Kutta method and applied to the scalar field via operator-splitting. Numerical effects of the scalar field can potentially lead to fractional abundances larger than the equilibrium value; therefore, for each cell we calculate the equilibrium value for the current density and prevent the fractional abundance from exceeding this value.

\subsubsection{Chemistry Tests}\label{sss:chemtests}

\begin{figure*}
  \includegraphics[width=\textwidth]{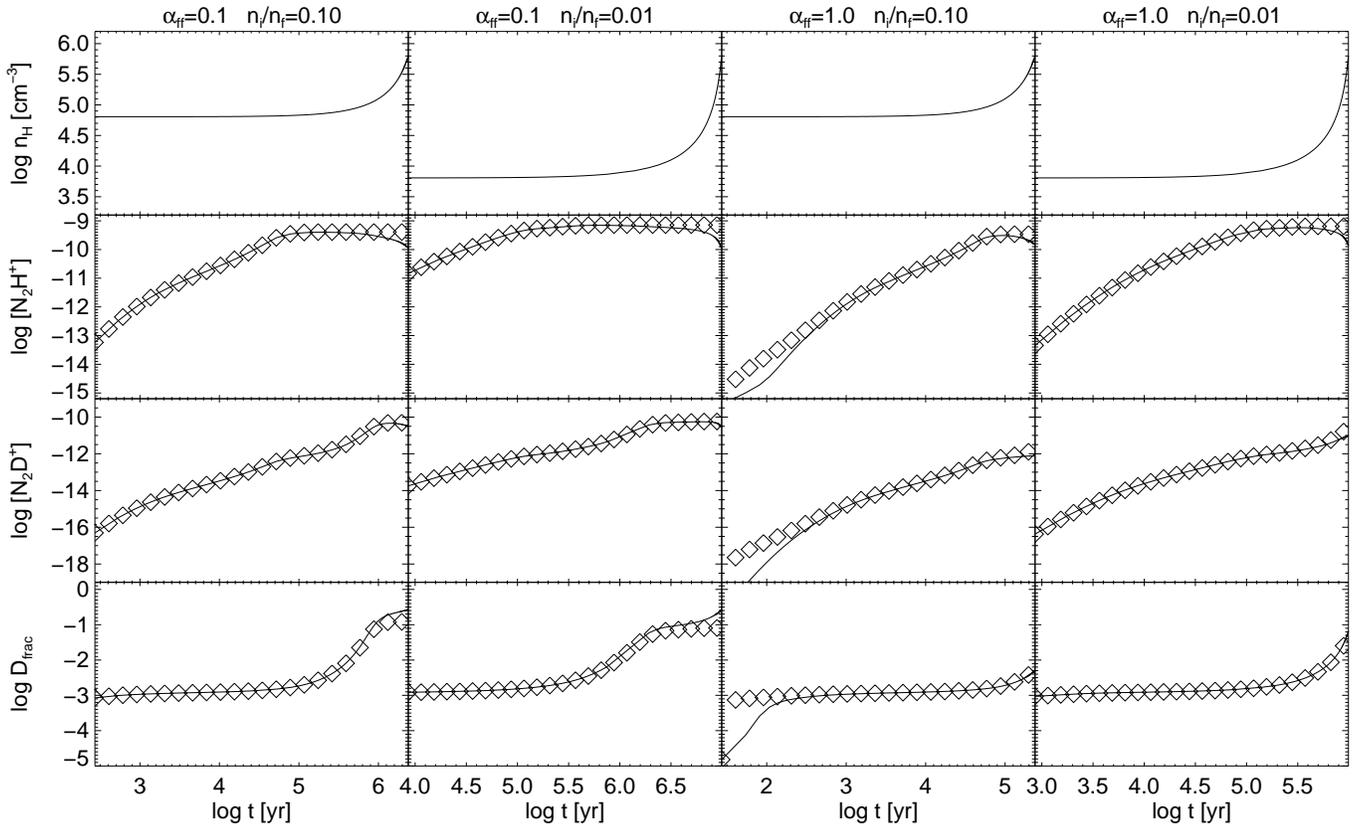}
  \caption{Comparison between K15 chemical network calculations (solid line) and our approximate chemical model in \textsc{Athena} (open diamonds) for Dynamic Density Evolution (DDE) tests. Each column shows a unique test case, with varying rates of collapse ($\alpha_{\rm ff}$) and density ratios ($n_{\rm i}/n_{\rm f}$). The evolution of the density (top row) is identical in both K15 and \textsc{Athena}; therefore no comparison is shown. Results are for \OPR~=~0.1. Overall, the results agree to within 30\%, with the largest discrepancies at initialization, as the short chemical timescales are difficult to resolve. At late times, there is a small tendency to systematically overestimate \nthp, which can lead to an underestimation of \Dfrac due to our parameterization method.}
  \label{f:afftest}
\end{figure*}

We validate our approximate chemical model by comparing to results from K15. We first compare simulations run with constant density. Overall, we find our approximate chemistry matches the full network calculations to within a few percent. As these models form the basis for our approximate method, it is reassuring that we match the evolution of all quantities accurately. We note that our parameterization leads to a systematic underestimate of the equilibrium values of [\nthp] and \Dfrac{}, up to 30\% below the values of K15. As discussed above, we need to make the growth rate a single-valued function of the current abundance, so we remove the slight decrease in [\nthp] near equilibrium.

We next compare to the Dynamic Density Evolution (DDE) simulations of K15. In these models, the authors used a single zone in which the hydrogen number density $n_{\rm H}$ evolved as
\begin{equation}
\frac{d n_{\rm H}}{dt} = \alpha_{\rm ff} \frac{n_{\rm H}(t)}{t_{\rm ff}(t)},
\end{equation}where $t_{\rm ff}$ is the local free-fall time at the current density. Results are shown in Figure \ref{f:afftest}. Overall, the results agree to within 10\% for most of the simulations. At early times, the short chemical time-scales are not well-resolved. Again, at late times the inability of [\nthp] to decrease leads to a systematic underestimate of [\nthp] and \Dfrac.

\section{Results}\label{s:results}

\begin{figure*}
  \includegraphics[width=\textwidth]{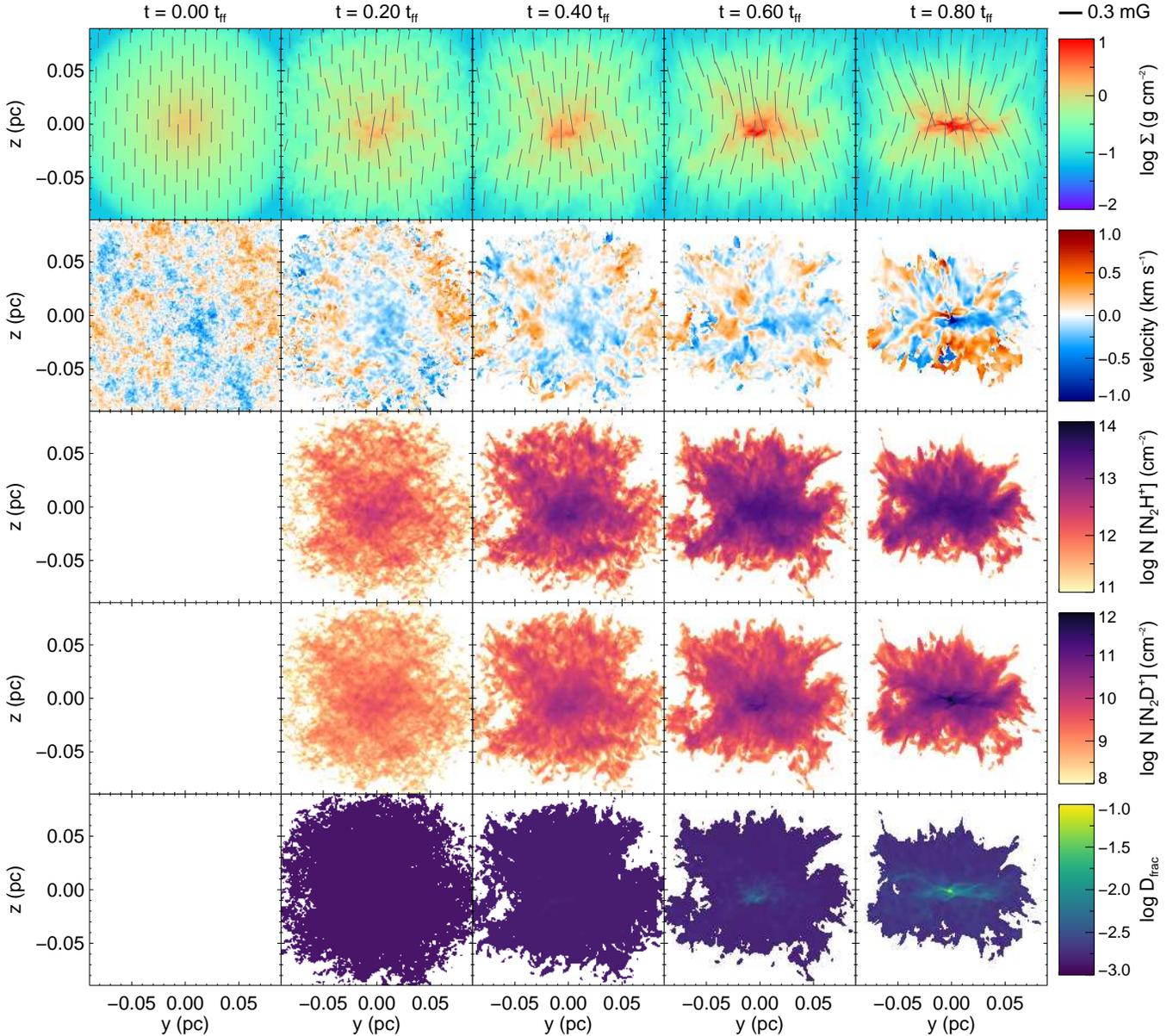}
  \caption{Projections from our fiducial turbulent, magnetized core model (run S3M2). Time proceeds from left to right in units of the initial mean free-fall time $t_{\rm ff}$. From top to bottom, the rows are: the mass surface density $\Sigma$; the mean velocity along the line of sight weighted by \ntdp; the column density of \nthp; the column density of \ntdp; and the deuterium fraction \Dfrac. The chemical starting time $t_{\rm chem} = 0~t_{\rm ff}$ and the initial ortho-to-para ratio of H$_2$ \OPR~=~0.1. Projections are taken along the $x$-axis, perpendicular to the initial magnetic field direction. The density-weighted magnetic field projection in the plane-of-sky is overlaid on the mass surface density in black lines, with the length proportional to the field strength. For reference, the length corresponding to $B = 0.3$~mG is shown in the top right. The chemical tracers are only considered where the molecular hydrogen number density is greater than $n_{\rm eff} = 4\times 10^5$~cm$^{-3}$, roughly 10\% of the critical density for the (3--2) transition. As [\ntdp]=0 at t=0, we instead show the density-weighted mean velocity for that panel only.}
  \label{f:S3M2timesnaps1}  
\end{figure*}

\begin{figure*}
  \includegraphics[width=\textwidth]{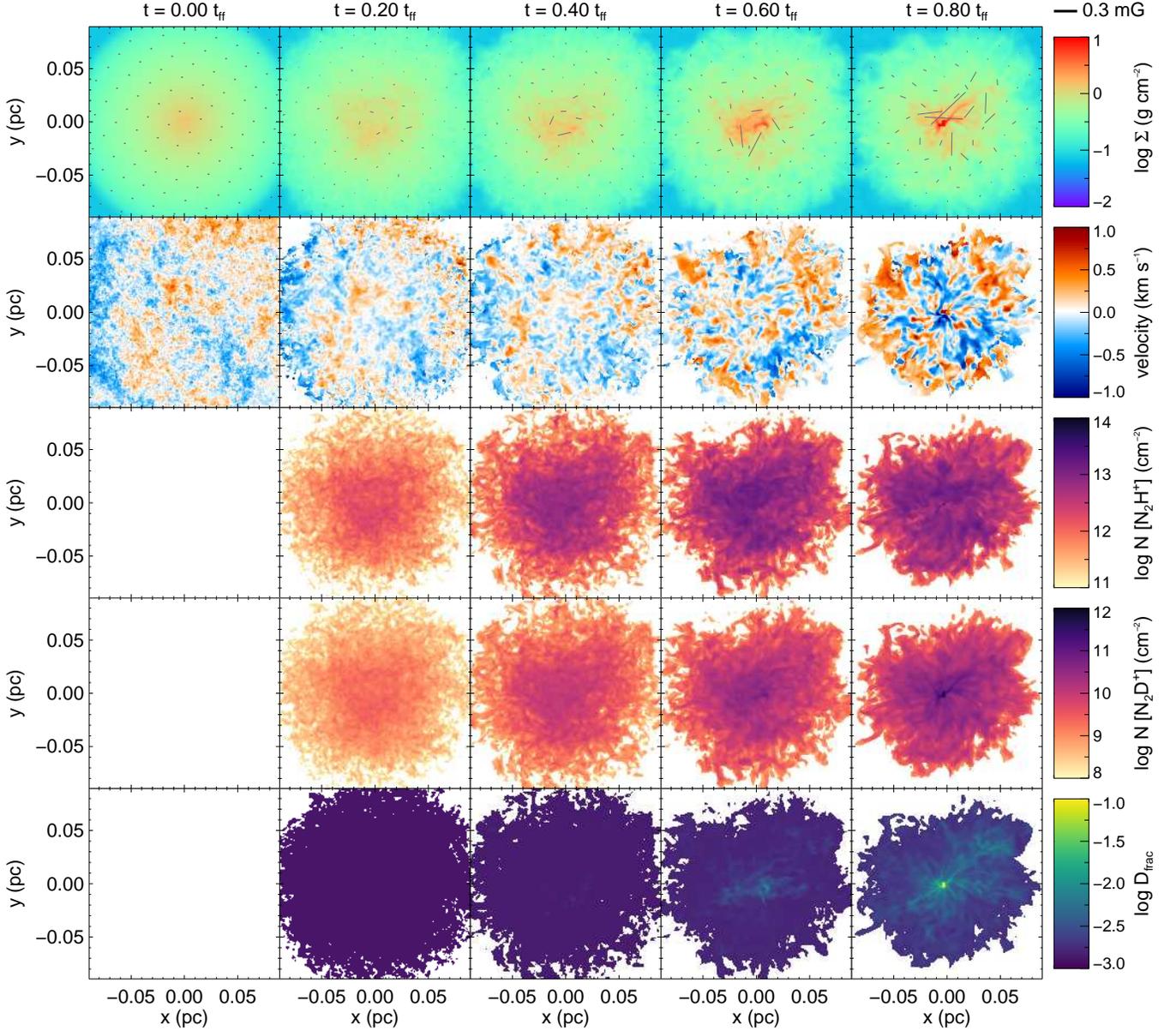}
  \caption{Same as Figure \ref{f:S3M2timesnaps1}, but now the projection is taken along the $z$-axis, parallel to the initial field direction.}
  \label{f:S3M2timesnaps3}
\end{figure*}

\begin{figure}
  \includegraphics[width=\columnwidth]{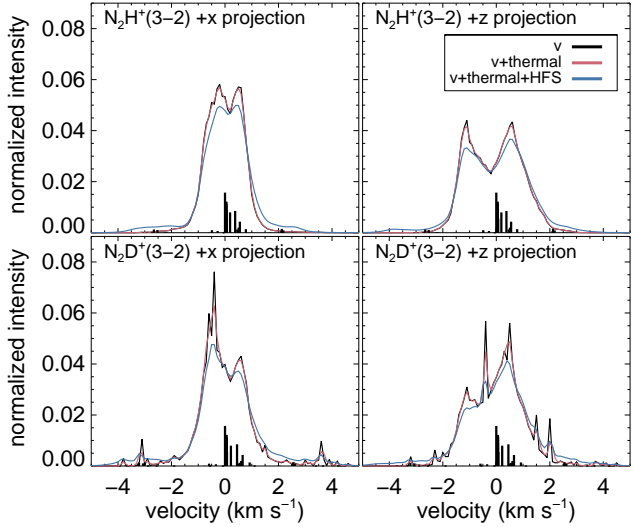}
  \caption{``Spectra'' of the total emission from \nthp(3--2) (top row) and \ntdp(3--2) (bottom row) in run S3M2 at simulation termination, for line-of-sight velocities parallel (left column) and perpendicular (right column) to the magnetic field. The velocity binsize is 0.1~km~s$^{-1}$. Assuming the gas is optically thin, we weight the velocities with the abundance of the tracer species. The black line shows the unprocessed distribution; the red line shows the effect of thermal broadening at $T=15$~K; and the blue line includes both thermal and hyperfine structure (HFS) broadening. For reference, the normalized HFS intensities are shown in black at the bottom of each panel.}
  \label{f:spectra}
\end{figure}

\begin{figure}
  \includegraphics[width=\columnwidth]{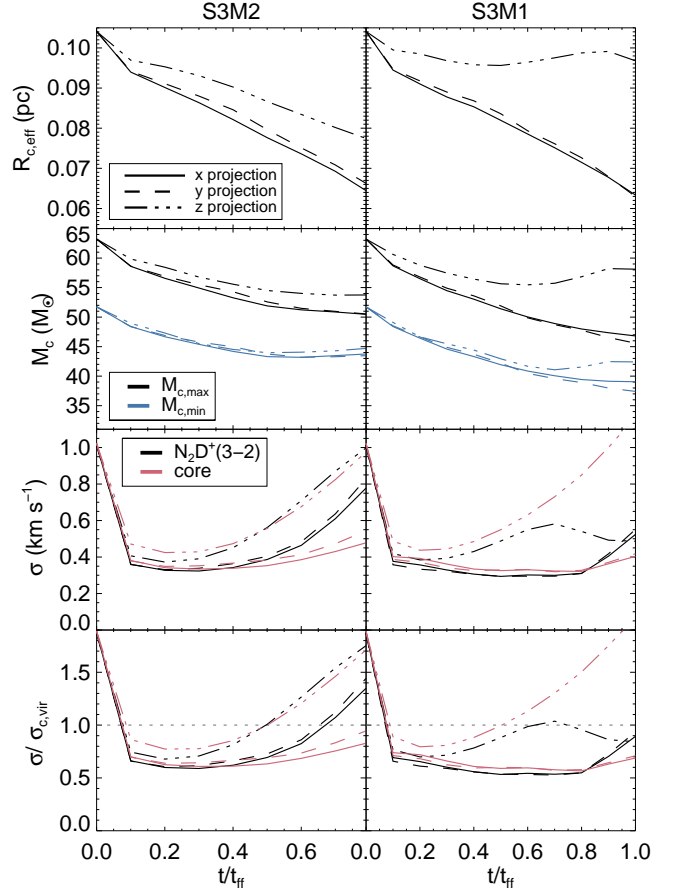}
  \caption{Time evolution of T13 observed quantities in runs S3M2 (left column) and S3M1 (right column). The rows show, from top to bottom: the effective core radius $R_{\rm c,eff}$; the core mass determined both with the clump contribution ($M_{\rm c,max}$; black) and without ($M_{\rm c,min}$; blue); the one-dimensional velocity dispersion of \ntdp(3--2) ($\sigma_{{\rm N}_2{\rm D}^+}$; black) and core tracer color field ($\sigma_{\rm C}$; red); and the ratio of $\sigma$ to the mass-averaged velocity dispersion of a virialized core $\sigma_{\rm c,vir}$, computed from Eq. \ref{eq:sigmacvir} using the minimum core mass (using $M_{\rm c,max}$ instead results in a 5\% increase in $\sigma_{\rm c,vir}$). $\sigma$ is determined from the total thermally-broadened spectra projected along the three Cartesian lines-of-sight (solid: x-direction; dashed: y-direction; dash-dotted: z-direction). For reference, $\sigma/\sigma_{\rm c,vir} = 1$ is indicated with a dotted horizonatal line. As [\ntdp] = 0 at $t=0$, we show instead the total velocity dispersion for this data point only.}
  \label{f:sigmaevo}
\end{figure}

\begin{figure}
  \includegraphics[width=\columnwidth]{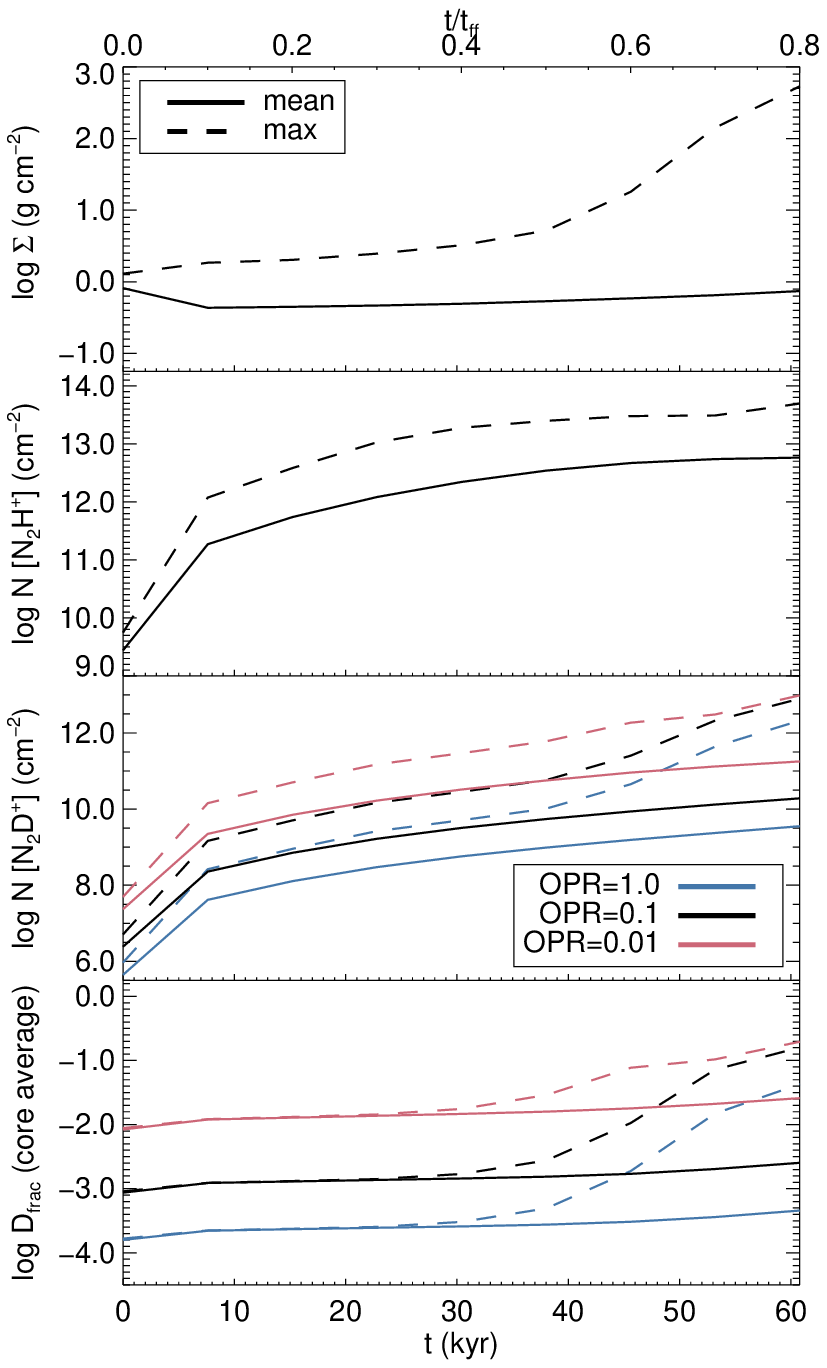}
  \caption{Time evolution of the mean (solid lines) and maximum (dashed lines) values of mass surface density $\Sigma$, \nthp{} column density, \ntdp{} column density, and deuterium fraction \Dfrac{} in the fiducial core (run S3M2). For \ntdp{} and \Dfrac, results are presented for different initial ortho-to-para ratio of H$_2$ {(blue: \OPR~=~1.0; black: \OPR~=~0.1; red: \OPR~=~0.01)}. Time is given in units of the initial core averaged free-fall time $t_{\rm ff}$ (top x-axis) as well in absolute time (bottom x-axis). The lower the initial \OPRH, the faster the deuteration proceeds. By the end of the simulation, the only estimate for mean \Dfrac{} that is similar to observations ($\gtrsim 0.1$) is for \OPR~=~0.01.}
  \label{f:meanevo}
\end{figure}

\begin{figure}
  \includegraphics[width=\columnwidth]{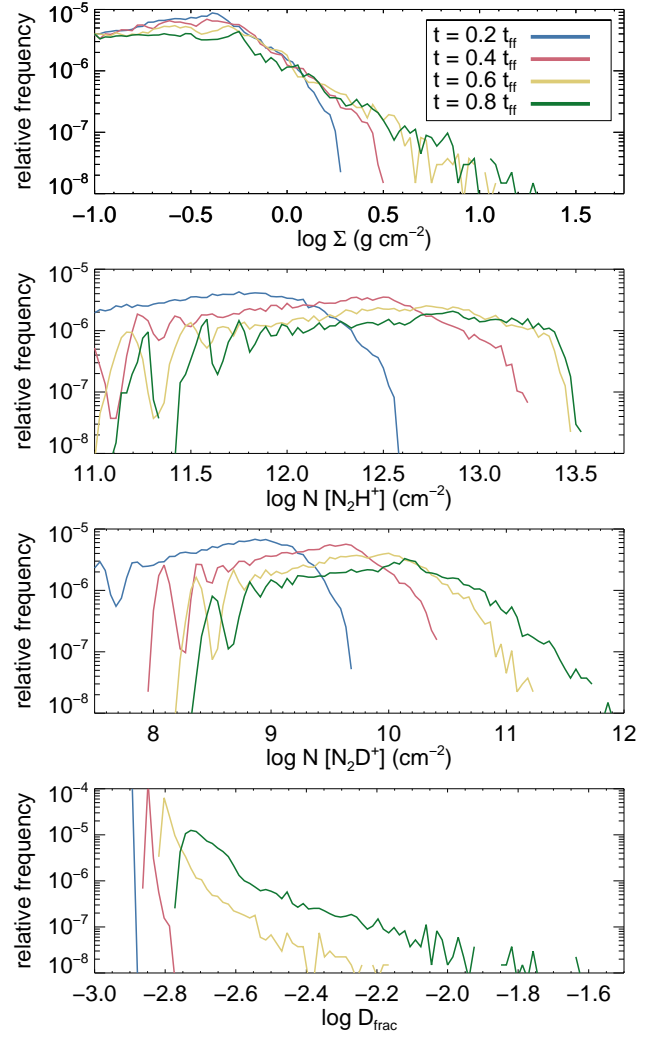}
  \caption{Probability distribution functions in our fiducial simulation (run S3M2) at multiple times. From top to bottom, the panels show the mass surface density $\Sigma$, the \nthp{} column density, the \ntdp{} column density, and the deuterium fraction \Dfrac. Simulation times are indicated by color (blue: $t=$0.2~$t_{\rm ff}$; red: 0.4~$t_{\rm ff}$; yellow: 0.6~$t_{\rm ff}$; green: 0.8~$t_{\rm ff}$). Projections are taken along the $x$-axis for $t_{\rm chem} = 0$ and \OPR~=~0.1.}
  \label{f:colpdftime}  
\end{figure}

\begin{figure}
  \includegraphics[width=\columnwidth]{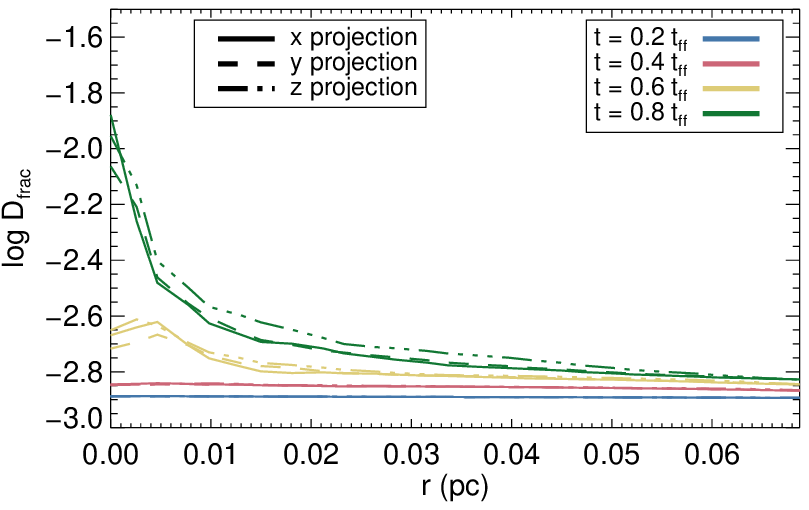}
  \caption{Radial averages of \Dfrac{} in our fiducial simulation (run S3M2) for $t_{\rm chem} = 0$ and \OPR~=~0.1. Results are shown for different times (blue: 0.2~$t_{\rm ff}$; red: 0.4~$t_{\rm ff}$; yellow: 0.6~$t_{\rm ff}$; green: 0.8~$t_{\rm ff}$) and projection directions (solid: $x$-axis; dashed: $y$-axis; dash-dotted: $z$-axis).}
  \label{f:radavgtime}
\end{figure}

\subsection{Dynamical Evolution}\label{ss:dynamicalevo}

We follow the collapse and chemical evolution of the fiducial simulation (run S3M2) for $ 61~{\rm kyr} \approx 0.8 t_{\rm ff}$. Figure \ref{f:S3M2timesnaps1} shows the time evolution of the mass surface density, mean velocity (weighted by the \ntdp{} abundance), and chemical tracers projected along the $x$-axis, perpendicular to the initial magnetic field orientation, as well as the ratio of the column densities (\Dfrac{} $\equiv N[$\ntdp$]/N[$\nthp$]$). For comparison to observations, we apply a density threshold when calculating the \nthp{} and \ntdp{} column densities based on the $J=3$--$2$ transition critical densities, which are given in table 2 of \citet{2013A&A...555A..41M}. For simplicity, we use a single value for both species of $n_{\rm crit}(3$--$2) \approx 4 \times 10^6~{\rm cm}^{-3}$. However, emission still occurs at densities below $n_{\rm crit}$ \citep{1999ARA&A..37..311E}, resulting in an effective critical density roughly an order of magnitude lower \citep{2015PASP..127..299S}; we therefore consider contributions to the chemical column densities only where $n_{\rm H_2} \ge n_{\rm eff} = 4 \times 10^5~{\rm cm}^{-3}$. The density-weighted plane-of-sky magnetic field projection is overlaid on the mass surface density map. 

The initial turbulent velocity field disrupts the smooth density distribution. The external pressure prevents significant expansion, and the core begins to collapse due to gravity. As these are ideal MHD simulations, the magnetic field in the $z$-direction prevents significant collapse along the perpendicular directions due to flux-freezing. Material can freely collapse along the field lines, creating an elongated filamentary structure in the $x$-$y$ plane. We follow the evolution of the core until runaway gravitational collapse into a few central cells prevents further evolution; this is essentially the formation of the first protostar. As seen in Figure \ref{f:S3M2timesnaps1}, the core collapses monolithically with little fragmentation, and the density appears to be centrally concentrated at termination. The magnetic field structure eventually develops an hourglass morphology as the field lines are pulled inward at the midplane. 

The asymmetry introduced by the magnetic field suggests the viewing angle will be important. Figure \ref{f:S3M2timesnaps3} shows projections taken along the $z$-axis, parallel to the initial field orientation. The core now appears circular, suggesting a disk-like structure in the $x$-$y$ plane. More small-scale structure is visible, as the velocity perturbations tangle and amplify the plane-of-sky magnetic field in the core; however, the central condensation remains distinct, surrounded by less-dense filaments or streams.

In the mean velocity map at $t=0.8 t_{\rm ff}$ in Figure \ref{f:S3M2timesnaps1}, there is a velocity gradient of several km~s$^{-1}$ across the central condensation, suggesting rotation in the $x$-$y$ plane. This is further evidenced in the velocity ``spectra'' shown in Figure \ref{f:spectra}, which all exhibit a double-peaked distribution. The spectra are computed from the integrated intensity of \nthp(3--2) and \ntdp(3--2) assuming LTE optically-thin emission ($j_X \propto n[X]$). To examine the effect of different broadening mechanisms, Figure \ref{f:spectra} presents the abundance-weighted velocity distribution with no broadening (black line), with thermal broadening at $T=15$~K (red line), and with hyperfine structure (HFS) broadening (where each component has the same Gaussian profile with a thermal velocity dispersion corresponding to $T=15$~K; blue line). The thermal velocity dispersion is sufficiently small ($\sigma_X \approx 0.06$~km~s$^{-1}$) that the broadening has only a modest effect. The projection direction has a pronounced effect, as the dispersion in the $+z$-direction (parallel to the magnetic field) is much wider than in the $+x$-direction. This may be attributed to material collapsing freely along the magnetic field lines. There are also noticeable differences between the \nthp{} and \ntdp{} spectra; the \ntdp{} spectra exhibit more small scale structures than the \nthp{}. As will be discussed in \S\ref{ss:chemicalevo}, \nthp{} largely reaches equilibrium throughout the core, whereas \ntdp{} does not; \ntdp{} may therefore probe smaller and denser structures within the core.

T13 assessed the virial state of observed pre-stellar cores by comparing the velocity dispersion of \ntdp(3--2), $\sigma_{{\rm N}_2{\rm D}^+}$, to the predictions for a virialized core based on MT03, $\sigma_{\rm c,vir}$ (Eq. \ref{eq:sigmacvir}). We present a similar analysis in Figure \ref{f:sigmaevo}. At each time step, we calculate the projected area in which \ntdp(3--2) emission is present, then ascribe an equivalent area circle to determine the effective core radius $R_{\rm c,eff}$. The clump mass surface density $\Sigma_{\rm cl}$ is then determined within the annulus from $R_{\rm c}$ to $2 R_{\rm c}$. To match the observations of T13, the effective core mass is determined from the projections in two ways: 1) the total mass surface density $\Sigma$ is summed within the equivalent area to compute the maximum core mass $M_{\rm c,max}$; 2) the clump surface density is subtracted from the total mass surface density before summing to compute the minimum core mass $M_{\rm c,min}$, removing contributions from the foreground and background to the core mass. The velocity dispersion of \ntdp(3--2) is determined by fitting a Gaussian to the thermally-broadened spectrum computed along each Cartesian projection direction. For comparison, we also show the velocity dispersion calculated from the core color tracer ($\sigma_{\rm c}$), which should represent the actual velocity dispersion of the core. Finally, the clump mass surface density and minimum core mass are used to estimate the velocity dispersion of a virialized core, $\sigma_{\rm c,vir}$ (Eq. \ref{eq:sigmacvir}), compared to $\sigma_{{\rm N}_2{\rm D}^+}$ and $\sigma_{\rm c}$.

In the fiducial run, the effective radius decreases as the core collapses. The effective core mass also decreases due to the central concentration of the \ntdp{} tracer. The core is initialized with a velocity dispersion $\sigma = 1~{\rm km~s}^{-1}$; yet by $t=0.1 t_{\rm ff}$, $\sigma \approx 0.4~{\rm km~s}^{-1}$. The velocity dispersion then increases most strongly in the $z$-direction, as material collapses freely along the magnetic field lines. As \ntdp becomes concentrated in the densest regions of the core, it no longer traces the overall velocity distribution and diverges from the color field estimate. The cores analyzed in both T13 and \citet{2016arXiv160906008K} were determined to be moderately sub-virial, with $\sigma_{{\rm N}_2{\rm D}^+}/\sigma_{\rm c,vir} \sim 0.8$ (based on the mm continuum estimate of core mass, which is expected to already include subtraction of the clump mass surface density via interferometric spatial filtering and thus be consistent with using $M_{c,{\rm min}}$). However, for the case of the massive core C1-S, T13 found $\sigma_{{\rm N}_2{\rm D}^+} / \sigma_{\rm c,vir} \simeq 0.45$ and argued this may imply the presence of strong ($\sim 1$~mG), large-scale magnetic fields. Here, we observe that after the initial turbulent energy injection, the fiducial core (run S3M2) appears moderately sub-virial but later becomes super-virial as the core collapses. The simulation with a stronger magnetic field (run S3M1), which is discussed in more detail in \S\ref{ss:mageffect}, shows an even more sub-virial velocity dispersion when viewed in the $x$ and $y$ directions, consistent with the T13 estimate for C1-S.

The evolution of the fiducial run is further quantified in Figure \ref{f:meanevo}, which shows the evolution of both the mean and maximum values of the mass surface density, chemical abundances, and \Dfrac{} in the core. Here we define the core using the effective number density threshold $n_{\rm eff}$; as this selects a unique volume, the mean column density is independent of viewing angle. Maximum values are computed from the $x$-axis projections. The mean mass surface density of the core decreases initially due to the initial turbulence and then increases slowly with time, from $\Sigma \approx 0.4$~g~cm$^{-2}$ up to 0.8~g~cm$^{-2}$. The maximum value increases nearly two orders of magnitude between 0.5 and 0.7 $t_{\rm ff}$, as the central overdensity contracts rapidly. The chemical evolution is discussed in \S\ref{ss:chemicalevo}. 

The same density threshold is applied to the column density probability distribution functions (PDFs) presented in Figure \ref{f:colpdftime}. As the core collapses, the initially (roughly) lognormal mass surface density distribution develops a high-density power-law tail, indicative of collapse. At simulation termination, roughly 10\% of the core mass is at $\Sigma \ge 1.0$~g~cm$^{-2}$.

\subsection{Chemical Evolution}\label{ss:chemicalevo}

As the density increases due to gravitational collapse, the growth rates of the chemical species also increase. We observe in Figures \ref{f:S3M2timesnaps1} and \ref{f:S3M2timesnaps3} that \nthp{} reaches equilibrium before \ntdp{} and is more widespread. This agrees well with observations of pre-stellar core regions; K16 find an extended envelope of \nthp{} emission around cores in IRDC G028.37+00.07, while \ntdp{} is more concentrated. The asymmetry introduced by the magnetic field also affects the chemical morphology; when viewing perpendicular to the field, the chemical tracers are more centrally-concentrated. The chemical evolution is also quantified in Figure \ref{f:meanevo}. The mean \nthp{} column density increases rapidly and then flattens over time as equilibrium is reached; in contrast, the mean \ntdp{} column density grows steadily throughout the simulation without reaching equilibrium, and \Dfrac{} increases only modestly until late times (after \nthp{} has reached equilibrium). The maximum values of \ntdp{} and \Dfrac{} do reach equilibrium values, but this is limited to only the densest regions of the core. This is confirmed in Figure \ref{f:colpdftime}, which shows only a small fraction of cells in the core are able to reach \Dfrac$ \ge 0.1$ by the end of the simulation. As K16 detected widespread deuteration in pre-stellar cores (see also the study of \citet{2016MNRAS.458.1990B} for evidence of widespread deuteration on parsec-sized, lower-density scales in an IRDC), this suggests more time is needed for the outer regions of the core to reach observed values.

Figure \ref{f:radavgtime} presents radial averages of \Dfrac{} within the core at different times and projection directions. \Dfrac{} grows rapidly in the center of the core, where the density is highest, while in the outer regions, \Dfrac remains relatively unchanged for the duration of the simulation. The direction of projection does not significantly affect the radial profile, which suggests observed radial profiles could be a useful (viewing-angle-independent) means to constrain the age of the core. Radial mapping of \Dfrac{} within observed cores is now technically feasible with ALMA.

\subsection{Effect of Initial \OPRH}\label{ss:opreffect}

\citet{2013A&A...551A..38P} and K15 found that the initial ortho-to-para ratio of H$_2$ (\OPR) strongly affected the chemical evolution of \ntdp{} and \Dfrac. Our fiducial simulation has \OPR~=~0.1. As the hydrodynamics is unaffected by the chemistry, we simultaneously evolve the molecular species using \OPR~=~0.01 and \OPR~=~1.0. The evolution of the mean values of \ntdp{} and \Dfrac{} at different \OPR{} is shown in Figure \ref{f:meanevo}. As noted by K15, a lower \OPR{} leads to faster growth of \ntdp, as well as a larger equilibrium value of [\ntdp]. Since \nthp{} is unchanged by \OPR, \Dfrac{} also grows faster and reaches a higher value. The mean value of \Dfrac{} remains below the observed values ($\gtrsim 0.1$) even at the lowest \OPR{} (=0.01), indicating a longer core lifetime and/or earlier deuteration (see \S\ref{ss:tchemeffect}) is necessary. The effect of varying \OPR{} is also presented in Figure \ref{f:dfraccompare}, which shows the ratio of chemical column densities (i.e., \Dfrac) at the end of the fiducial simulation for varying \OPR. From the top row moving down, \OPR{} decreases for a given chemical age, with a corresponding increase in the mean deuterium fraction in the core. 

\subsection{Effect of Initial Chemical Age}\label{ss:tchemeffect}

\begin{figure*}
  \includegraphics[width=\textwidth]{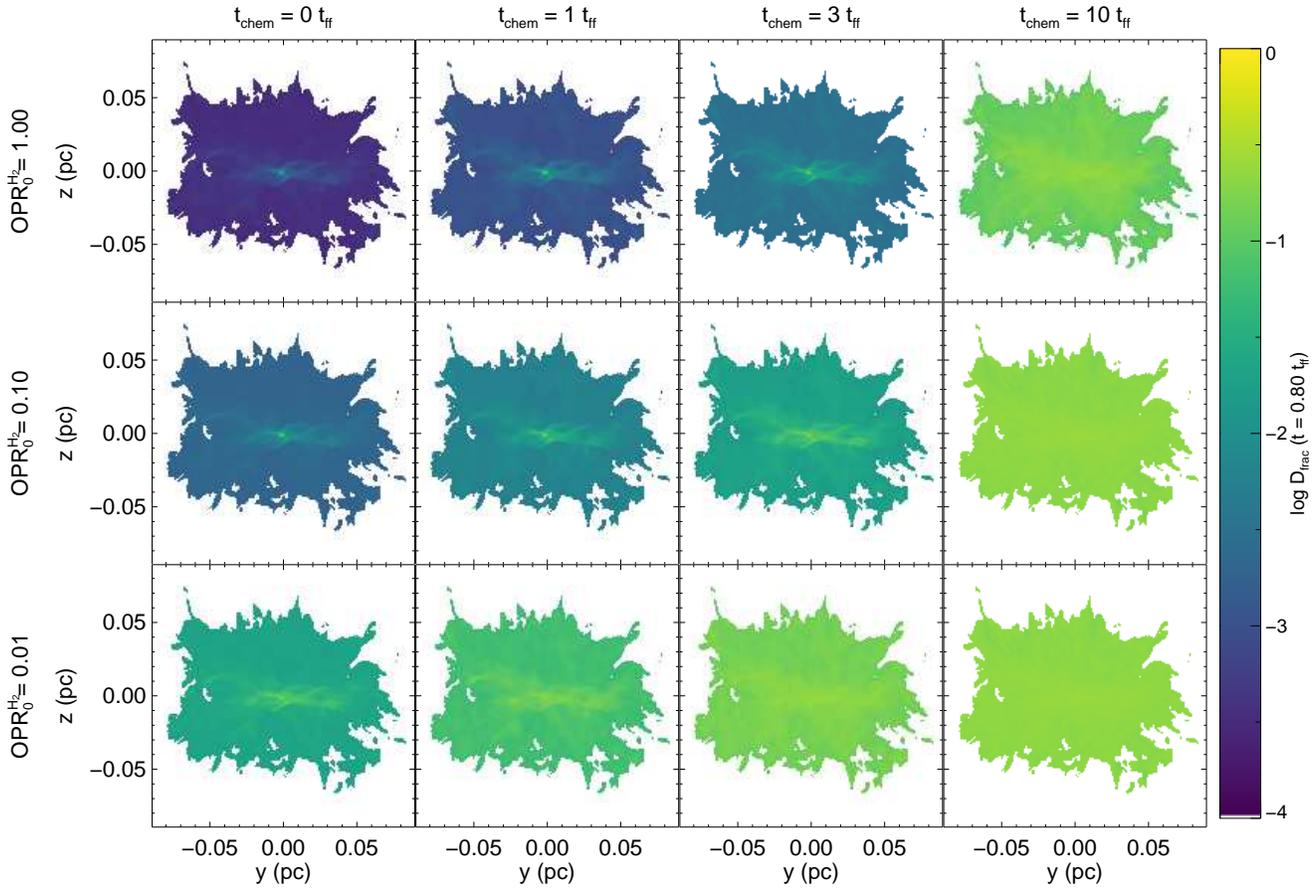}
  \caption{Ratio of chemical column densities (\Dfrac) at simulation termination ($t = 0.8 t_{\rm ff}$) from our fiducial model (run S3M2) for different initial chemical ages and ortho-to-para ratios of H$_2$. From left to right, the columns are at $t_{\rm chem} = $ 0, 1, 3, and 10 $t_{\rm ff}$; from top to bottom, the rows are \OPR~=~1.00, 0.10, and 0.01. As either $t_{\rm chem}$ or \OPR{} are increased, the resulting mean deuterium fraction in the core increases.}
  \label{f:dfraccompare}  
\end{figure*}

We have thus far assumed in our calculations that the gas begins in an initially pristine condition, with $t_{\rm chem} = 0$, i.e., [\nthp]=[\ntdp]=0.0 at $t=0$. However, this may not be the case, especially given our initial condition; we initialize the core after it has already formed a centrally-concentrated structure. Deuteration begins once CO begins to freeze-out, which occurred at some unknown time prior to the current state. We therefore explore different ``chemical ages'' for the core: $t_{\rm chem} =$ 0, 1, 3, and 10~$t_{\rm ff}$. To set the initial condition for the chemical abundances, we reference the constant density results of K15 at an absolute time, as described in \S\ref{ss:chemistry}. The core then begins from an advanced state of deuteration, assuming the core has been in its current density configuration for $t_{\rm chem}$. While the dynamical collapse is unchanged, the core is able to reach higher deuterium fractions. As is evident in Figure \ref{f:dfraccompare}, the deuterium fraction increases for increasing chemical age, with nearly the entire core achieving the equilibrium value of \Dfrac{} for $t_{\rm chem} = 10 t_{\rm ff}$. While this may seem to agree with the estimates of K15, which indicated up to 10 free-fall times may be necessary to reach observed values of \Dfrac, the simulations are not directly comparable. In K15, the density continually increases, with a corresponding decrease in $t_{\rm ff}$; here, we assume a constant density (hence a constant $t_{\rm ff}$) prior to initialization. Regardless, in both cases the conclusion remains that deuteration must proceed for longer than the average free-fall time, either by earlier deuteration or slower collapse.

\subsection{Effect of Initial Mass Surface Density}\label{ss:sigmaeffect}

\begin{figure*}
  \includegraphics[width=\textwidth]{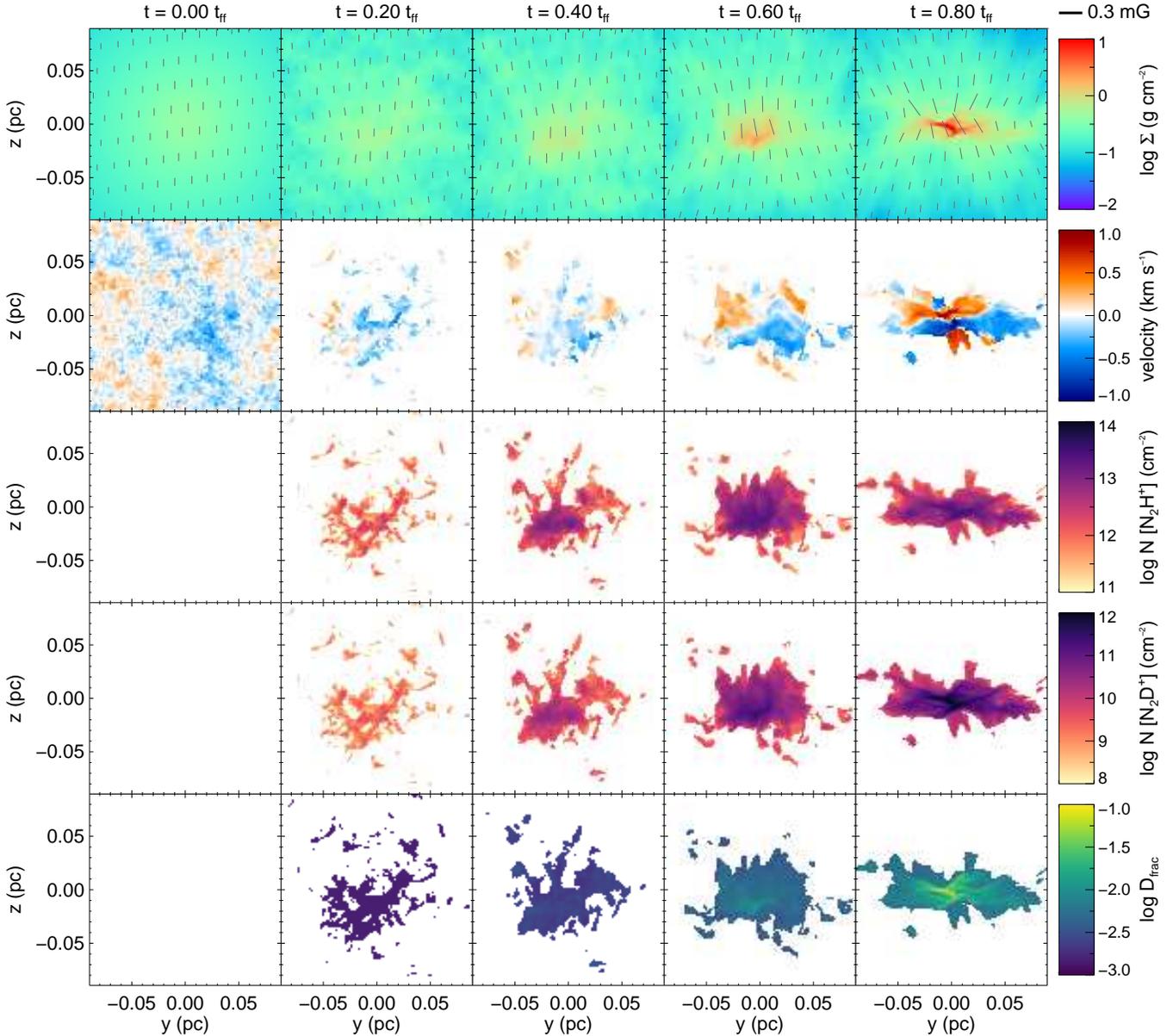}
  \caption{Similar to Figure \ref{f:S3M2timesnaps1} but for a core with lower initial mass surface density ($\Sigma_{\rm cl} = 0.1$~g~cm$^{-2}$; run S1M2). Projections are taken along the $x$-axis (perpendicular to initial magnetic field direction). The simulation runs to the same relative time ($0.8~t_{\rm ff}$), which corresponds to a longer absolute time (139~kyr).}
  \label{f:S1M2timesnaps1}
\end{figure*}

\begin{figure*}
  \includegraphics[width=\textwidth]{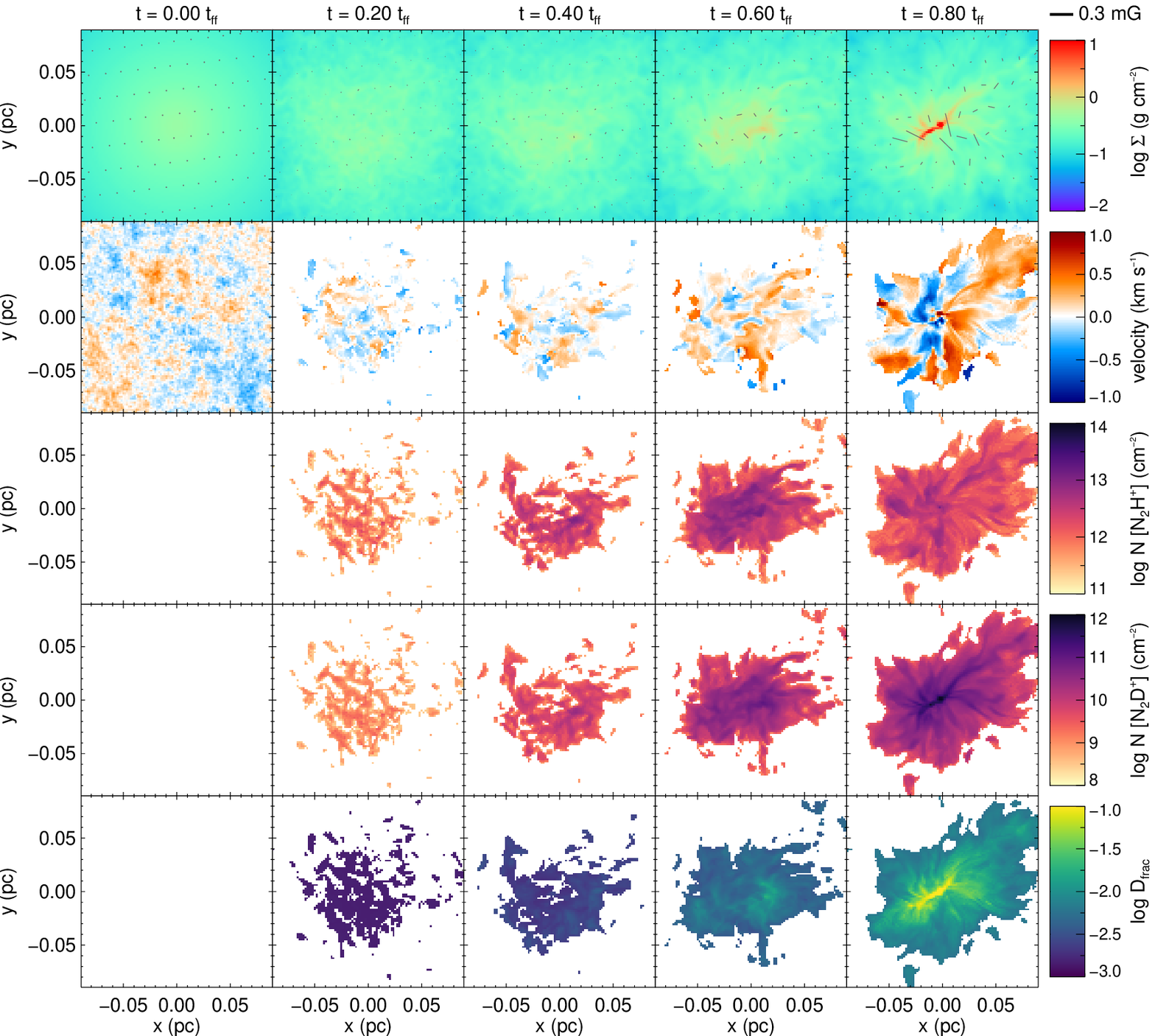}
  \caption{Same simulation as in Figure \ref{f:S1M2timesnaps1} (run S1M2) but now the projections are taken along the $z$-axis, parallel to the initial magnetic field direction.}
  \label{f:S1M2timesnaps3}
\end{figure*}

\begin{figure}
  \includegraphics[width=\columnwidth]{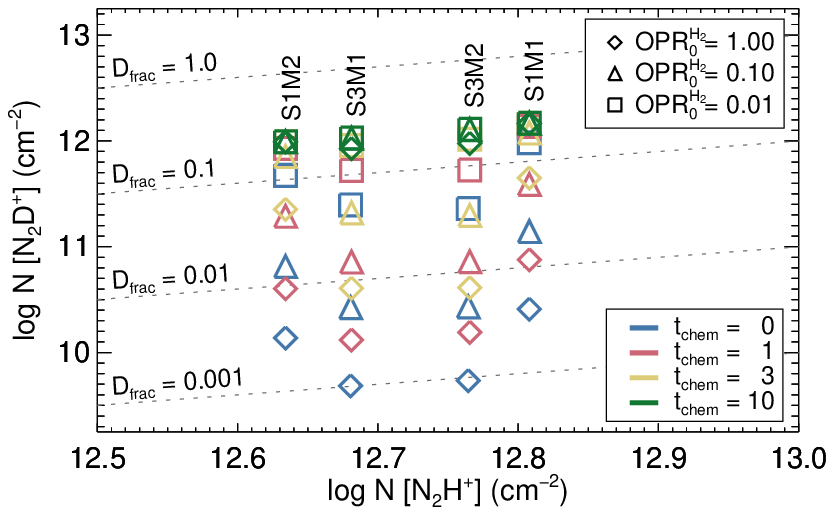}
  \caption{Summary of mean chemical column densities in the core for all runs at simulation termination, for varying initial chemical age (indicated by color) and ortho-to-para ratio of H$_2$ (indicated by symbol). Reference lines for \Dfrac{} are indicated with dashed lines. The \nthp{} column density is largely constant across the parameter space of each run because 1) \nthp{} is largely unaffected by changes in \OPRH; and 2) equilibrium is reached for all values of $t_{\rm chem}$. Values of \Dfrac$ \ge 0.1$ are only reached for low values of \OPR{} or large chemical age.}
  \label{f:results}
\end{figure}

We also examine the effect of varying the initial clump mass surface density $\Sigma_{\rm cl}$. Our fiducial simulation uses $\Sigma_{\rm cl} = 0.3$~g~cm$^{-2}$; however, this is the current observed state of the cores in the T13 sample. As the cores currently show significant deuteration, we investigate an earlier phase of the core lifetime by decreasing the initial clump mass surface density to $\Sigma_{\rm cl} = 0.1$~g~cm$^{-2}$ (run S1M2). We keep the core mass fixed at 60~M$_\sun$ and use the prescription of MT03 to adjust the core radius (increase $R_{\rm c} \to 0.18$~pc) and surface number density (decrease $n_{{\rm H},s} \to 8.2 \times 10^4$~cm$^{-3}$). The average core free-fall time then increases to $t_{\rm ff} \to 173$~kyr. We maintain the core temperature at $T_{\rm c}=15$~K and the initial virial parameter $\alpha = 2$; the initial velocity dispersion then decreases to $\sigma \to 0.76 $~km~s$^{-1}$. We also maintain the same mass-to-flux ratio $\mu_\Phi = 2$; the central field strength is then reduced to $B_{\rm c} \to 0.27$~mG.

Figures \ref{f:S1M2timesnaps1} and \ref{f:S1M2timesnaps3} show the evolution of run S1M2 for projections along the $x$- and $z$-axes, respectively. Based on the results of K15 presented in Figure \ref{f:denscompare}, we expect the lower densities in the core to lead to slower chemical growth and lower equilibrium values of \Dfrac. The core collapses more slowly on an absolute timescale, but the simulation terminates at the same relative time, $t= 0.8~t_{\rm ff}$. Comparing on a relative timescale, there are only modest differences in morphology and chemistry between the two cases. At termination, run S1M2 appears more filamentary and less centrally concentrated than run S3M2. Although the absolute column density values are lower, the mean \Dfrac{} is actually higher. This is displayed in Figure \ref{f:results}, which shows the final mean chemical column densities and corresponding \Dfrac{} for all simulations performed. Depending on the value of $t_{\rm chem}$ and \OPR, \Dfrac{} is higher in run S1M2 by a factor of 1-5 over run S3M2. We therefore conclude that the initial mass surface density does not strongly affect the chemical evolution.

\subsection{Effect of Magnetic Field Strength}\label{ss:mageffect}

\begin{figure*}
  \includegraphics[width=\textwidth]{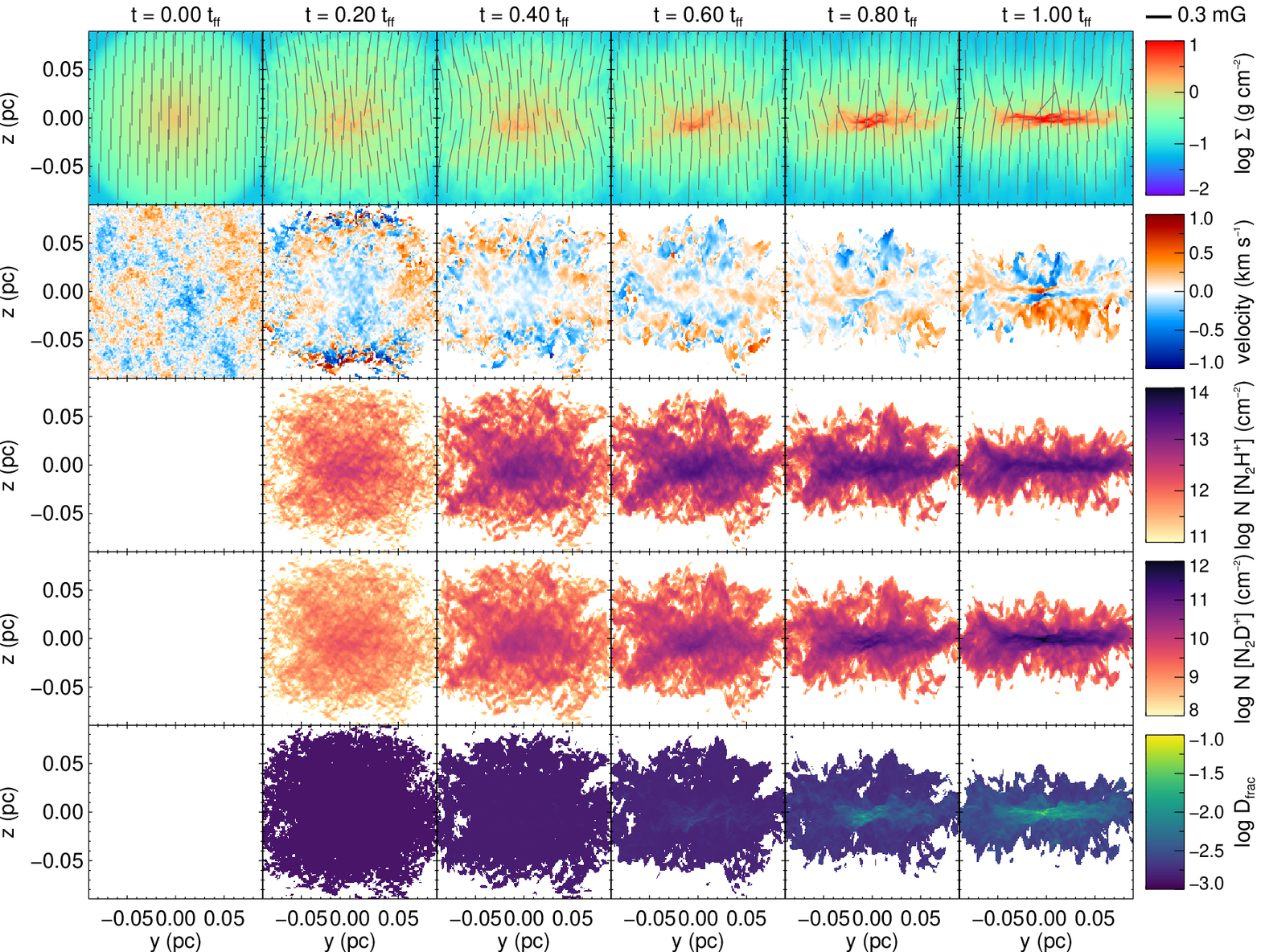}
  \caption{Similar to Figure \ref{f:S3M2timesnaps1} but for a core with a stronger magnetic field ($\mu_{\Phi} = 1$; run S3M1). Projections are taken along the $x$-axis, perpendicular to initial magnetic field direction. The critical magnetic field inhibits the collapse, allowing the simulation to proceed another $0.2~t_{\rm ff}$. Due to flux-freezing, material collapses most freely parallel to the magnetic field; hence the core becomes compressed in the $z$-direction.}
  \label{f:S3M1timesnaps1}
\end{figure*}

\begin{figure*}
  \includegraphics[width=\textwidth]{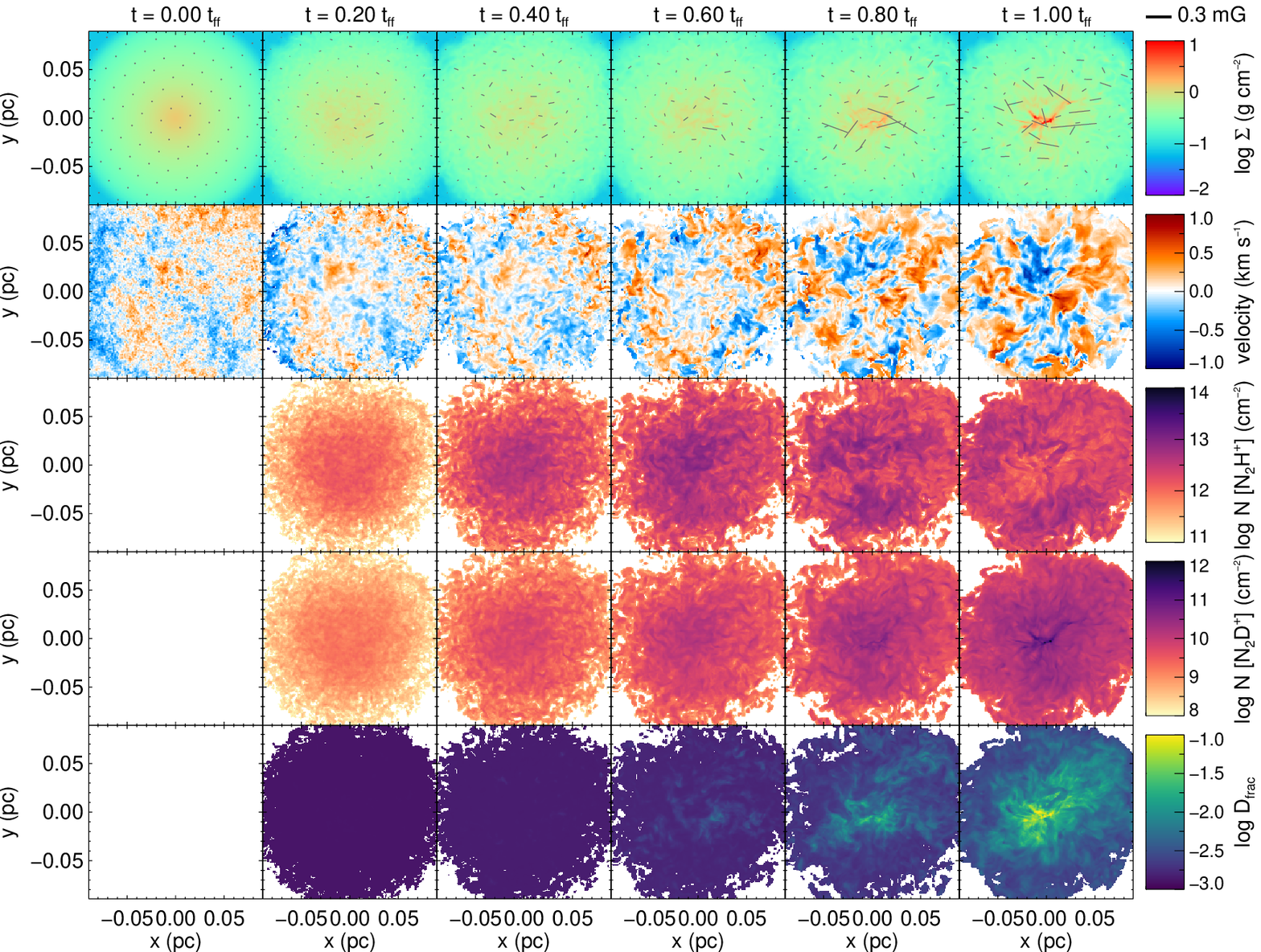}
  \caption{Same simulation as Figure \ref{f:S3M1timesnaps1} (run S3M1) but now the projections are taken along the $z$-axis, parallel to the initial magnetic field direction.}
  \label{f:S3M1timesnaps3}
\end{figure*}

\begin{figure*}
  \includegraphics[width=\textwidth]{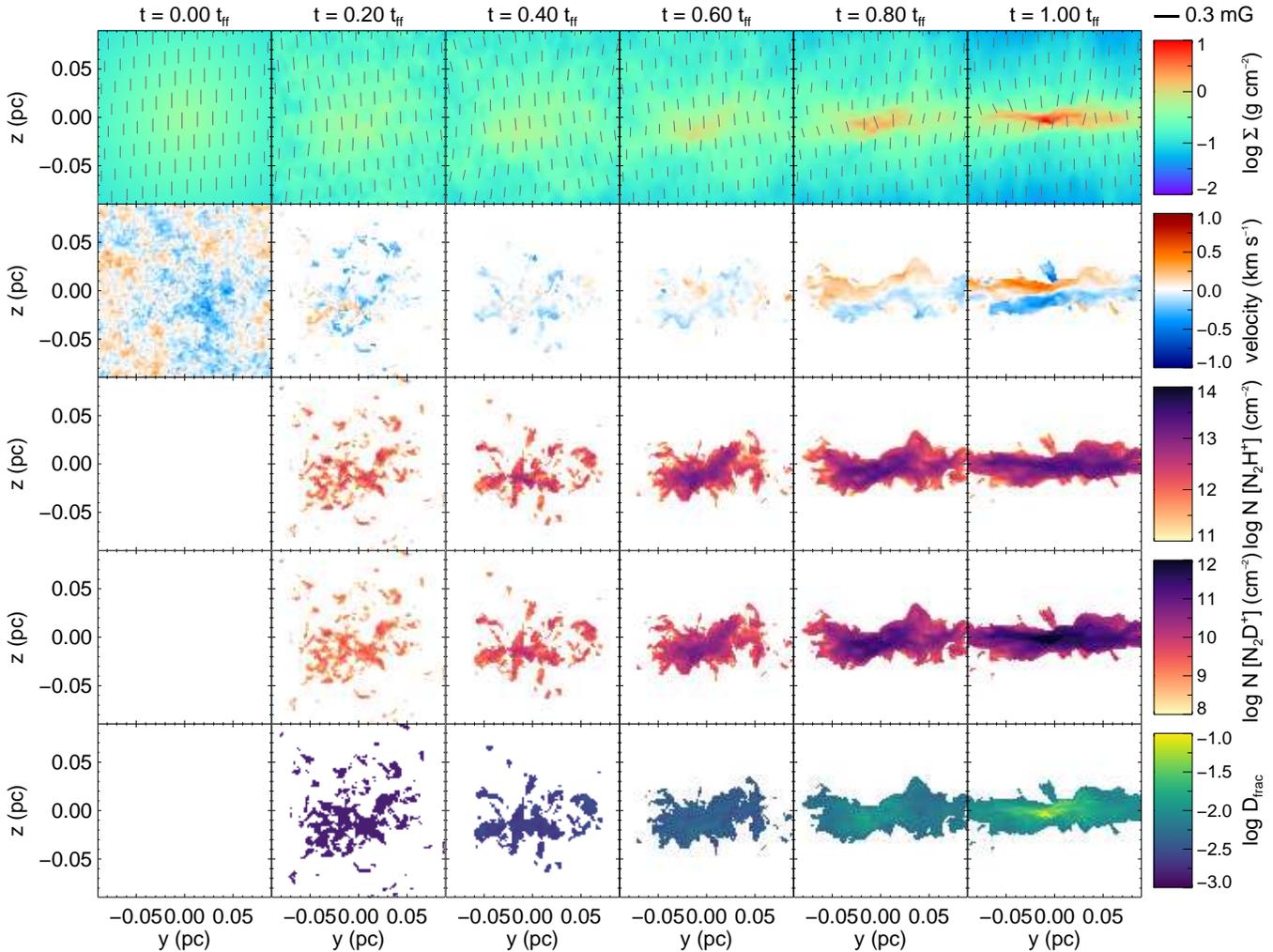}
  \caption{Similar to Figure \ref{f:S1M2timesnaps1} but for a core with increased magnetic field strength ($\Sigma_{\rm cl} = 0.1$, $\mu_{\Phi} = 1$; run S1M1). Projections are taken along the $x$-axis (perpendicular to initial magnetic field direction). As in Figure \ref{f:S3M1timesnaps1}, the stronger field again slows the collapse and leads to an elongated core.}
  \label{f:S1M1timesnaps1}
\end{figure*}

\begin{figure*}
  \includegraphics[width=\textwidth]{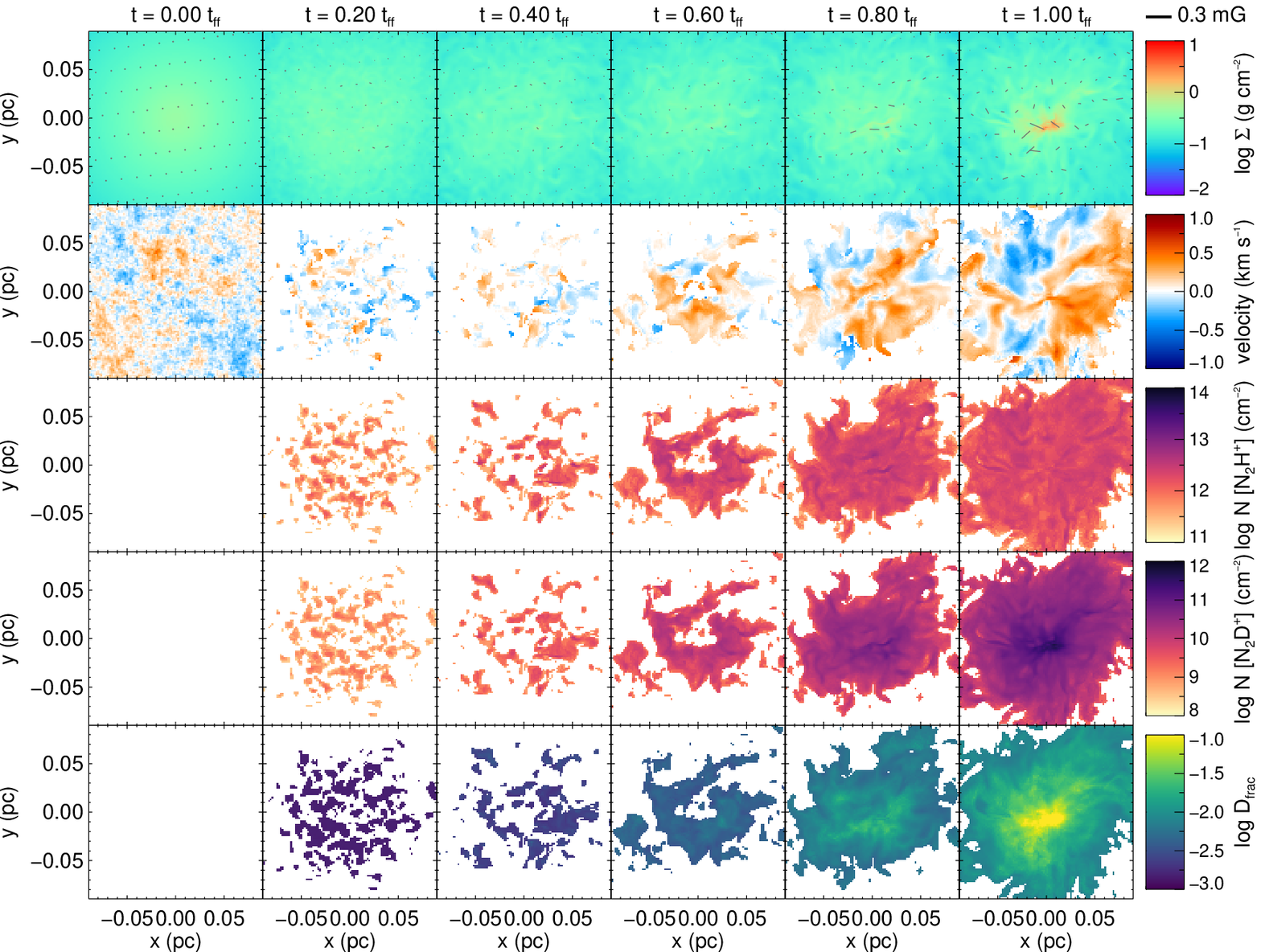}
  \caption{Same simulation as Figure \ref{f:S1M1timesnaps1} (run S1M1) but for projections taken along the $z$-axis, parallel to the initial magnetic field direction.}
  \label{f:S1M1timesnaps3}
\end{figure*}

We also test the effect of increasing the magnetic field strength. We perform simulations with a critical field strength ($\mu_\Phi = 1.0$) for both the fiducial mass surface density ($\Sigma_{\rm cl} = 0.3$~g~cm$^{-2}$; run S3M1) and the decreased value ($\Sigma_{\rm cl} = 0.1$~g~cm$^{-2}$; run S1M1). Projections are shown for run S3M1 in Figures \ref{f:S3M1timesnaps1} and \ref{f:S3M1timesnaps3}, and for run S1M1 in Figures \ref{f:S1M1timesnaps1} and \ref{f:S1M1timesnaps3}. In both instances, the stronger magnetic field leads to an initial expansion of the core before it coalesces again and collapses. The critical field does slow the contraction -- both simulations run $0.2t_{\rm ff}$ past the corresponding $\mu_\Phi=2.0$ simulations -- but ultimately does not prevent collapse. The slower collapse leads to a larger, more diffuse core compared to the fiducial run at a given time. The stronger field also inhibits motions perpendicular to the field, as illustrated in Figure \ref{f:sigmaevo}. The velocity dispersions in the $x$ and $y$ directions are lower in run S3M1 compared to run S3M2, while the $z$ direction is largely unaffected. The filamentary structure observed perpendicular to the field is also narrower, which reduces estimates of the mass.

Figure \ref{f:results} reveals that the longer timescale at $\mu_\Phi = 1.0$ does result in a higher mean value of \Dfrac{} in both cases, but only by a factor of 1-2 over runs with $\mu_\Phi =2.0$. As with the mass surface density, we conclude that the magnetic field strength does not strongly affect the resulting deuterium fraction. However, we caution that this result may be influenced by the initial field geometry (see \S\ref{s:discussion}), and further investigation is warranted.

\section{Discussion}\label{s:discussion}

Figure \ref{f:results} summarizes the final mean chemical column densities (and \Dfrac) for all simulations performed and across the entire parameter space. In agreement with the one-zone models of K15, we find that deuteration proceeds slowly during collapse and only reaches observed values under certain conditions, namely low \OPR{} ($\lesssim 0.01$) or advanced chemical evolution ($t_{\rm chem} \gtrsim 3 t_{\rm ff}$). The initial mass surface density and magnetic field strength do not largely alter this conclusion.

Our approximate chemistry model for \nthp{} deuteration is constructed from the results of K15. The K15 chemical network calculations were performed with the same physical conditions (e.g., temperature, ionization rate, dust-to-gas ratio) except for the density and \OPR. As noted in \S\ref{ss:chemistry}, these two quantities play a large role in determining the deuteration and are therefore parameters of our model. However, varying any of the other K15 model parameters could shift the equilibrium value of \Dfrac{} by an order of magnitude, as is evident in fig. 5 of K15. In particular, increasing the initial heavy-element depletion factor $f_{\rm D}$ decreases the timescale for deuteration. This may explain the discrepancy between our work and the results of \citet{2013A&A...551A..38P}; we set $f_{\rm D} = 10$ and obtain results consistent with slow collapse, whereas \citet{2013A&A...551A..38P} chose $f_{\rm D} \gtrsim 300$ and determined a fast collapse best matched \Dfrac observations. While there is evidence for some CO depletion ($f_{\rm D} \lesssim 5$) in IRDCs \citep{2011ApJ...738...11H,2012ApJ...756L..13H}, further observations are necessary to better constrain this parameter. Fig. 5 of K15 also demonstrates the effect of changing the cosmic ray ionization rate $\zeta$; increasing $\zeta$ will both decrease the equilibrium value of \Dfrac and increase the growth rate of \Dfrac. We also note that the K15 fiducial parameter values may not be applicable across the full range of densities ($10^3 \le n_{\rm H} \le 10^9 {\rm cm}^{-3}$). Short of implementing the full network in 3D MHD simulations, future work could introduce density-dependent parameters to better span the K15 parameter space. Finally, our parameterization introduces a systematic overestimate of \nthp{} and underestimate of \Dfrac{}, by up to 30\% relative error. While the approximation is expedient for simulations, a time-dependent chemical network will be required to obtain more accurate species evolution as the density evolves. 

The cores begin with a smooth density profile, and we rely on the initial super-virial turbulent velocity field to create density fluctuations. The turbulence is thus not fully developed at initialization, and the energy decays rapidly as seen in Figure \ref{f:sigmaevo}. Ideally, the density and velocity structure would be generated in a self-consistent manner, possibly through driven turbulence with subsequent application of gravity \citep[e.g.,][]{2001ApJ...547..280H,2014MNRAS.439.3420M}.

We consider magnetic fields but neglect non-ideal MHD effects, such as ambipolar diffusion (AD). The AD timescale is estimated in figure 6 of K15 to be roughly an order of magnitude longer than the free-fall timescale for all relevant densities \citep[see also ][]{2014MNRAS.443..230H}. As all our simulations terminate prior to $t=2~t_{\rm ff}$, we do not expect AD to significantly affect the dynamics. However, the field geometry may affect the results. We begin with a smooth, cylindrically-symmetric field in the $z$-direction. As with the density field, the magnetic field is tangled by turbulent motions, but only after initialization. Material freely collapses along the field lines even when the field is of critical strength. Future studies should begin either with a tangled component in addition to an ordered component, or should generate a tangled field through turbulent driving \citep{2014MNRAS.439.3420M}.

We are limited in the range of spatial scales we can probe due to the lack of mesh refinement. Our fiducial simulation is performed using a fixed grid of $512^3$ grid cells, for a minimum resolution of $\delta \approx 360$~AU. Following the collapse and chemical evolution further will require additional resolution, possibly through the use of mesh refinement. 

We also halt our calculations when the collapsing core is no longer adequately resolved, i.e., at protostar formation. We do not include sink particles, as we are only interested in pre-stellar conditions. For similar reasons, we also neglect radiative feedback. As demonstrated by \citet{2011ApJ...742L...9C} and \citet{2013ApJ...766...97M}, including radiation feedback from protostars slows the collapse and inhibits fragmentation. It is unclear how the protostellar radiation field will affect the deuteration process; however, there is recent evidence that protostars can exist within highly-deuterated regions \citep{2016ApJ...821L...3T}. Radiation effects could increase the core lifetime and hence the deuterium fraction, and future studies following the chemistry for longer periods should include these effects using sink particles and radiation-magnetohydrodynamics.

\section{Conclusions}\label{s:conclusions}
We have constructed an approximate chemical model for the deuteration of \nthp{} in cold, dense pre-stellar gas. Our model is based on the results of the astrochemical network presented in K15. The full network is prohibitively expensive in multi-dimensional hydrodynamics simulations. Rather than reducing the number of reactions, we parameterize the results across a range of densities into look-up tables. This approximate formulation is demonstrated to perform reasonably well in comparison to full network calculations with both constant and evolving density. 

We implement our approximate chemical model in the \textsc{Athena} MHD code. In 3D simulations, we follow deuteration during the collapse of a turbulent, magnetized pre-stellar core. The core is initialized in accordance with the Turbulent Core Accretion model of MT03. For our adopted initial conditions, the core collapses to the point of forming a protostar within roughly one free-fall time, regardless of the initial mass surface density or magnetic field strength. During most of this collapse phase the velocity dispersion of the core as traced by \ntdp(3--2) appears moderately sub-virial compared to predictions of the MT03 Turbulent Core Model, consistent with observations of T13 and \citet{2016arXiv160906008K}. Only near the end, just before protostar formation, does the velocity dispersion rise to appear super-virial. 

As the core collapses, the increase in density accelerates the deuteration of \nthp. However, we find that \Dfrac{} does not reach observed values ($\gtrsim 0.1$) in $\sim 1~t_{\rm ff}$, unless the initial ortho-to-para ratio of H$_2$ (\OPR) is $\lesssim 0.01$ or the core begins from an advanced chemical state ($t_{\rm chem} \gtrsim 3~t_{\rm ff}$). This is in agreement with K15 and suggests that the collapse rate in highly-deuterated cores may be significantly slower than the free-fall time, or the deuteration process begins earlier than assumed.

\acknowledgments{
We thank the anonymous referee for a helpful report. Computations were performed on the Kure cluster at UNC-Chapel Hill. MDG and FH gratefully acknowledge support by NC Space Grant and NSF Grant AST-1109085. SK and JCT acknowledge support from NSF Grant AST-1411527. PC acknowledges support from the European Research Council (ERC; project PALs 320620).}

\software{Athena \citep{2008ApJS..178..137S}}

\end{document}